\documentclass[a4paper,10pt,reqno]{amsart}
\usepackage[margin=2.5cm]{geometry}
\usepackage{graphicx}
\usepackage{dsfont,amsmath,amssymb,subfigure,array,rotating,natbib}

\newcommand\rod [1][d]{\rho^{ \scriptscriptstyle (#1)}} %\newcommand{\rod }{\rho^{ \scriptscriptstyle ({d})}}
\def\rhoB  {\rho^{ \scriptscriptstyle (\text{B})}} % 

\renewcommand\ij{i\!j}
  
  \newcommand{\vev   }[1]{\langle #1 \rangle}
  \newcommand{\Vev   }[1]{\left\langle #1 \right\rangle}
  \newcommand{\esp   }[1]{\mathds{E}[#1]}
  \newcommand{\var   }[1]{\mathds{V}[#1]}
  \newcommand{\pr    }[1]{\mathds{P}[#1]}
  \newcommand{\1     }[1]{\mathds{1}_{\{#1\}}}
  \newcommand{\Esp   }[1]{\mathds{E}\!\left[#1\right]}

  \newcommand{\Exp   }[1]{      \exp\!\left(#1\right)}
\def \be  {\begin{equation}}
\def \ee  {\end{equation}}

  \newcommand{\cdf }[1][]{\mathcal{P}_{\!\negthickspace{\scriptscriptstyle <}{,#1}}}
  \newcommand{\qdf }[1][]{\mathcal{P}_{\!\negthickspace{\scriptscriptstyle <}{,#1}}^{-1}} %{\mathcal{Q}_{#1}}

  \newcommand{\vect}[1]{\boldsymbol{#1}} % essayer aussi avec \mathbf
  \newcommand{\mat }[1]{\mathrm{#1}}
  \newcommand{\cop }[2][]{C_{#1}#2}

\def\argmin{\operatornamewithlimits{arg\,min}}

\def\Wei{\beta}                      %\mat{X}
\def\ret{x}                          %r
\def\Ret{\mat{\MakeUppercase{\ret}}} %\mat{R}

\newcolumntype{g}{>{$}r<{$}}

\def\ff{\scriptstyle \text{f{}f}}
\def\rr{\scriptstyle \text{rr}}
\def\fr{\scriptstyle \text{fr}}

\begin{document}

\title{A nested factor model for non-linear dependences in stock returns}

\author[R.~Chicheprotiche]{R\'emy Chicheportiche}
\address{Capital~Fund~Management, 75\,007 Paris, France\\
         Chaire de finance quantitative, Ecole Centrale Paris 92\,295 Ch\^atenay-Malabry, France}
\author[J.-P.~Bouchaud]{Jean-Philippe Bouchaud}
\address{Capital~Fund~Management, 75\,007 Paris, France}

\maketitle

\begin{abstract}
The aim of our work is to propose a natural framework to account for all the empirically known properties of the multivariate distribution of stock returns.
We define and study a ``nested factor model'', where the linear factors part is standard, but where the log-volatility of the
linear factors {\it and} of the residuals are themselves endowed with a factor structure and residuals. We propose a calibration procedure to estimate 
these log-vol factors and the residuals. We find that whereas the number of relevant linear factors is relatively large ($10$ or more), only two or three log-vol
factors emerge in our analysis of the data. In fact, a minimal model where only {\it one} log-vol factor is considered is already very satisfactory, as it accurately reproduces 
the properties of bivariate copulas, in particular the dependence of the medial-point on the linear correlation coefficient, as reported in \cite{chicheportiche2012joint}. 
We have tested the ability of the model to predict Out-of-Sample the risk of non-linear portfolios, and found that it performs significantly better than other schemes. 
\end{abstract}
%%%%%%%%%%%%%%%%%%%%%%%%%%%%%%%%%%%%

% \def\Wei{\beta}                      %\mat{X}
% \def\ret{x}                          %r
% \def\Ret{\mat{\MakeUppercase{\ret}}} %\mat{R}
%%%%%%%%%%%%%%%%%%%%%%%%%%%%%%%%%%%%%%%%%%%%%%%%%%%%%%%%%%%%%%%%%%%%%%%%
\section{Introduction}
Dependences among financial assets or asset classes stand at the heart of 
modern portfolio selection theories.
Whatever the (concave) utility of an investor and its risk measure, 
diversification is profitable but optimal diversification is only reached 
if the underlying dependence structure is well understood.

For example, the well-known Markowitz theory \citep{markowitz1952portfolio,markowitz1959portfolio,bouchaud2003theory} of optimal portfolio design 
aims at finding the optimal weights $w_i$ to attribute to each stock of a pool. 
It assumes that stock returns are correlated random variables $\ret_i$, and that
the optimizing agent has a ``mean-variance'' quadratic utility function in the form 
$U(\vect{w})=\esp{\vect{\ret}\cdot\vect{w}}-\mu\var{\vect{\ret}\cdot\vect{w}}$, with a parameter $\mu$ controlling for risk-aversion level.
It hence relies on the linear covariance matrix $\rho=\esp{\vect{\ret}\vect{\ret}^\dagger}$
of the stock returns, and more importantly on its inverse $\rho^{-1}$.
Indeed, with no further constraints (budget, transaction costs, operational risk constraint, 
prohibition of short selling, etc.), the optimal weights are given by
\begin{equation}
    \vect{w}_\rho^*\propto {\rho^{-1}\vect{g}} \equiv \mat{V}\Lambda^{-1}\mat{V}^\dagger\vect{g}
\end{equation}
where $\vect{g}$ is the vector of gain targets for the assets in the basket, 
and $\rho=\mat{V}\Lambda\mat{V}^\dagger$ is the spectral decomposition of the covariance matrix with 
$\mat{V}$ being the square matrix of eigenvectors and $\Lambda$ the diagonal matrix of eigenvalues.

Empirical estimates of $\rho$ and its spectrum $\Lambda$ are typically very noisy, 
and cleaning schemes need to be applied before inversion if one wants to avoid 
artificially enhancing the weights of low-risk in-sample modes 
--- as made clear by the above expression of $\vect{w}_\rho^*$ in terms of $\Lambda$ ---
that turn into high-risk realized out-of-sample modes.

All this is by now fairly standard practice, and several cleaning schemes have been designed, 
in view of modeling either the signal (parametric models, factor models, Principal Components Analysis),
or the noise (RMT-based \cite{laloux1999noise, laloux2000random, ledoit2004well,potters2005financial,elkaroui2010high,bartz2012directional}), 
see also \cite{tumminello2007shrinkage,potters2009financial}.
 
However, it is now established that markets operate beyond the Gaussian, linear regime.
For one thing, individual stock returns are well known to be non-Gaussian, 
and moments beyond the mean and variance have gained considerable interest 
(e.g.\ the excess kurtosis, or low-moment estimates thereof).
But more importantly, stock returns are \emph{jointly} not Gaussian: 
the structure of dependence between pairs of stocks is not compatible with the Gaussian copula,
and as a consequence the penalty in the utility function should be more subtle than just
the portfolio variance and include non-linear measures of risk (like tail events, quadratic correlations, etc.) 
in order to better fit the agent's risk aversion profile.
Only in a multivariate Gaussian setting can these non-linear dependences be fully expressed in terms of the linear correlations.

\begin{figure}
    \center
    \includegraphics[scale=.6,trim=0 15 0 0,clip]{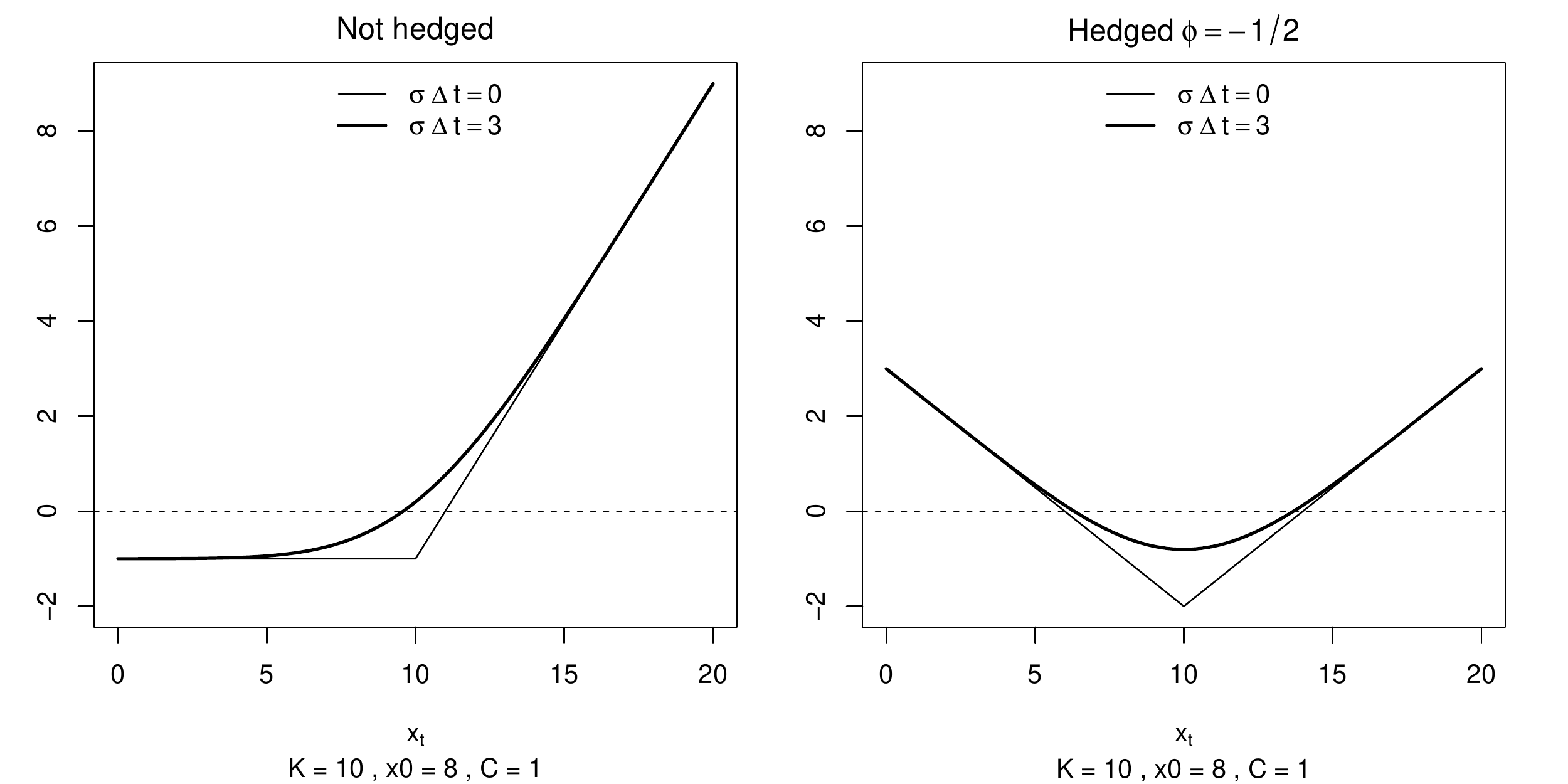}
    \caption{Expected payoff of an option as a function of the current price of the underlying stock.
             \textbf{Left:} unhedged; \textbf{Right:} hedged by short selling $|\phi|=1/2$ shares of the underlying.
             The illustration is for a call option of strike $x_K=10$, price $\mathcal{C}=1$ 
             on a stock of initial price $x_0=8$ following a Bachelier diffusion with volatility $\sigma$.
             The thin line is the payoff at expiry ($\sigma\,\Delta t=0$) and 
             the thick curve is the expected value of the payoff before expiry ($\sigma\,\Delta t=3$). }
    \label{fig:payoff_option}
\end{figure}

Non-linear dependences are also very important in the pricing and risk management of structured products and portfolios of derivatives.
For example, the payoff of a hedged option has a V-shape with linear asymptotes and quadratic core, see Fig.~\ref{fig:payoff_option}.
A portfolio of several such hedged options has thus a variance characterized by the absolute and quadratic correlations of the underlying stocks (gamma risk).
The correlations of these amplitudes, needed for estimating the risk at the portfolio level, are even noisier than linear correlations, 
whence the need for a reliable model of both linear and non-linear dependences.

In a previous article \citep{chicheportiche2012joint}, we showed that the joint distribution of daily returns of stocks is not ``elliptical'' either, 
which is to say that stocks are not exposed to a unique volatility volatility affecting all of them.
We in fact ruled out all models with a single stochastic volatility $\sigma$, of the form
\begin{equation}\label{eq:ellipt}
    \ret_i=\sigma\,\epsilon_i,
\end{equation}
with jointly Gaussian (and correlated) residuals $\epsilon_i$'s.
This, we argued, revealed a finer structure in the non-linear dependences,
and opened the way for a description taking into account several modes of volatility.
However, our results also showed that any description in the form of individual volatilities 
\begin{equation}\label{eq:pseudo_ellipt}
    \ret_i=\sigma_i\,\epsilon_i,
\end{equation}
with {\it arbitrary dependences} between the $\sigma_i$'s, would not be able either to explain successfully the empirical joint distribution. 
We focused in particular on the medial point of bivariate copulas, $\cop(\frac12,\frac12)$,%
\footnote{
The copula $\cop(u_i,u_j)$ of a random pair $(X_i,X_j)$ is the joint probability 
that the variables are below their marginal $(u_i,u_j)$-th quantiles respectively:
\[
    \cop(u_i,u_j)=\pr{\ret_i<\qdf[i](u_i),\ret_j<\qdf[j](u_j)},
\]
where $\cdf[i]$ denotes the cumulative distribution function (CDF) of $\ret_i$.}
which is the probability that both variables are below their median value simultaneously. 

\begin{figure}
          \center
          \subfigure[2000--2004]{\includegraphics[scale=.38,trim=0 0 0 25,clip]{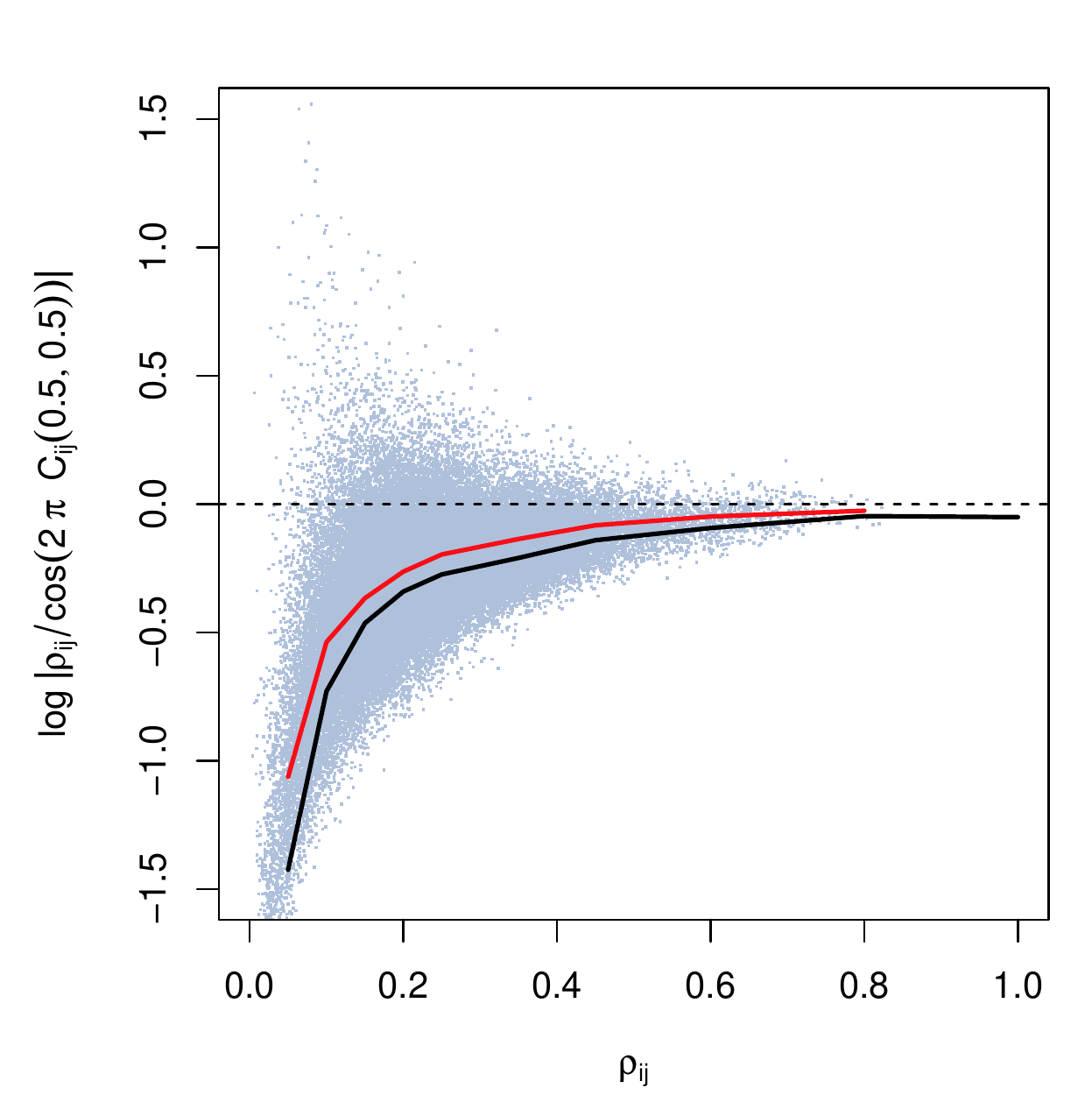}}
          \subfigure[2005--2009]{\includegraphics[scale=.38,trim=0 0 0 25,clip]{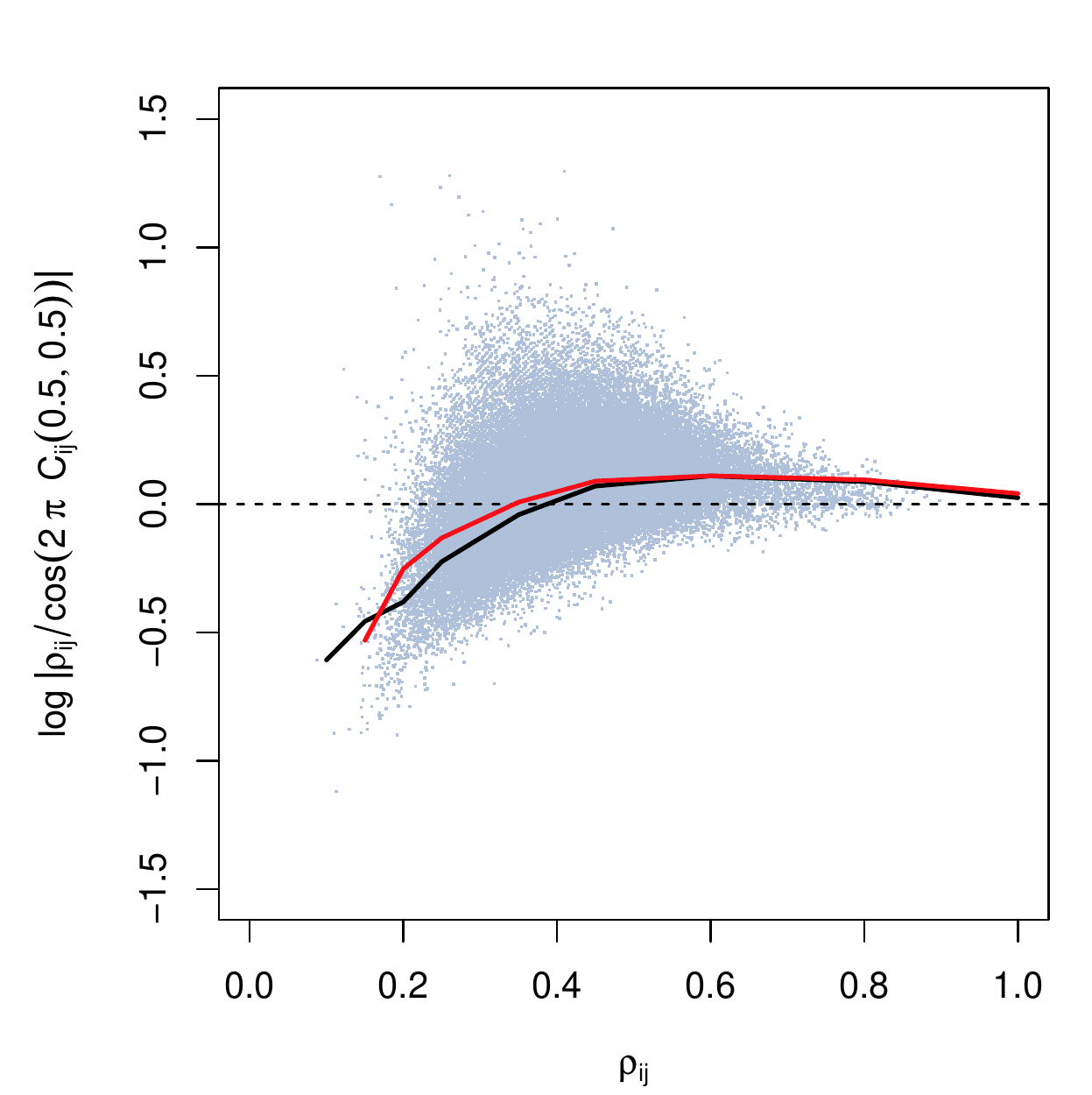}}
          \subfigure[2009--2012]{\includegraphics[scale=.38,trim=0 0 0 25,clip]{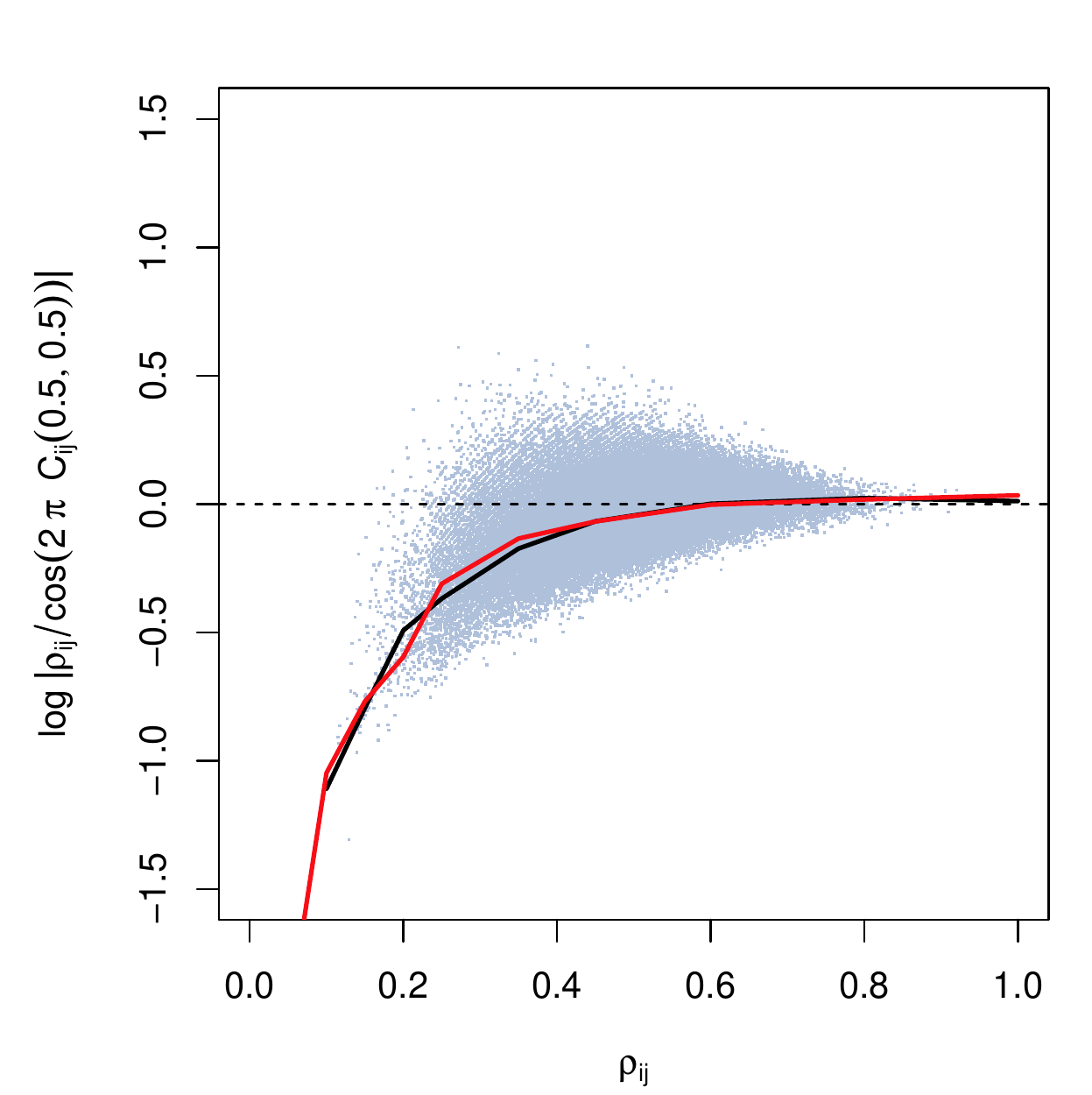}}
          \caption{$\ln|\rho/\rhoB|$ vs $\rho$ for each stock pair, see Eq.~\eqref{eq:hatrho}. 
                   This figure quantifies the departure of the medial point of the copula $\cop(\tfrac12,\tfrac12)$
                   from the pseudo-elliptical benchmark defined by Eq.~\ref{eq:pseudo_ellipt}, 
                   for which the prediction is a straight horizontal line at 0 (dashed).
                   Empirical values are shown as a scatter plot of all stock pairs (grey cloud) 
                   as well as bin averages (black line), for the 3 periods considered here.
                   The ``dominant volatility mode'' prediction for the average behavior (see Sect.~\ref{sec:modeling_vol}) 
                   is shown in red and agrees remarkably well with data, and captures correctly its time evolution. Note that
                   $\ln|\rho/\rhoB|$ tends to 0 when $\rho$ increases, which shows that highly correlated stocks are more
                   ``elliptical'', i.e.\ indeed exposed to the same volatility mode.}
          \label{fig:emp_cop}
          \end{figure}

All pseudo-elliptical models (defined by Eq. (\ref{eq:pseudo_ellipt})) 
lead to a simple relation between the medial copula to the coefficient of linear correlation, 
see the discussion in \cite{chicheportiche2012joint}:
\[ %begin{equation}
    \cop[ij](\tfrac12,\tfrac12)=\frac{1}{4}+\frac{1}{2\pi}\arcsin\,\rho_{ij}
\] %end{equation}
Said differently, the effective correlation%
\footnote{The superscript (B) stands for ``Blomqvist'', as $\rhoB$ is related to 
Blomqvist's beta coefficient, see \citep{blomqvist1950measure}.}
\begin{equation}\label{eq:hatrho}
    \rhoB_{ij}\equiv\cos\!\left(2\pi \cop[ij](\tfrac12,\tfrac12)\right)
\end{equation}
is equal to $\rho_{ij}$ for these models, whereas empirical data shows marked departures from this prediction, see \ Fig.~\ref{fig:emp_cop}.
As discussed in \cite{chicheportiche2012joint}, the scatter plot of $\ln|\rho/\rhoB|$ vs $\rho$ for every stock pair 
is not concentrated around the elliptical prediction
(dashed horizontal line crossing the $y$-axis at 0), but rather the average curve (black line) departs significantly from the prediction.
Furthermore, the less correlated the pairs are, the farther they depart from an elliptical bivariate distribution, 
calling for a richer description than just amplitudes exposed to a {\it common mode} of fluctuations.
         
This, together with a comparison of empirical and theoretical values of other observables (e.g.\ copula diagonals), motivates a 
description of stock returns with several modes of volatility, but which excludes models where the $\sigma_i$'s can be decomposed 
multiplicatively into a market contribution $\omega$, a sectorial contribution $\widehat\omega_s$ (where stock $i$ belongs to sector $s$), 
and a residual contribution $\widetilde\omega_i$ as:
\[
    \sigma_i=\sigma_0 e^{\omega + \widehat\omega_s + \widetilde\omega_i}.
\]
Instead, we proposed that additive non-Gaussian factors should be able to generate anomalous medial copula values,
because of the interplay of factor kurtosis and residual kurtosis,
as motivated by the toy model for $\cop(\frac12,\frac12)$ presented in \cite{chicheportiche2012joint}. 

The search for a theoretical description of multivariate dependences has led to the explosion of the literature on copulas. 
Several families of copulas have been proposed, beyond the elliptical one discussed above: Archimedean (Clayton, Franck, and others), Vine, Liouville, etc.
Unfortunately, as emphasized in \cite{chicheportiche2012joint}, these copulas are often theoretical figments with no financial 
interpretation, and for that reason alone should be considered with suspicion. (These alternative copulas also fail to 
reproduce the empirical dependences of stock returns). We advocated in \cite{chicheportiche2012joint} the need for 
constructing {\it meaningful} copulas, based on intuition and plausibility. The aim of the present paper is therefore
to construct a general factor model, flexible enough to reproduce all the
known stylized facts of the empirical joint distribution of stock returns, but still simple enough to 
be easily calibrated on data. We want our multivariate model of stock returns to be able to:
\begin{itemize}
    \item Reproduce the structure of linear correlations with a small number of factors. 
    \item Generate fat-tailed return series, with non-Gaussian factors and residuals;
    \item Allow for a dependence between the volatilities of the residuals and the volatilities of the factors, 
          as observed in  \cite{cizeau2001correlation,allez2011individual}.
    \item Reproduce the anomalous copula structure determined in \cite{chicheportiche2012joint}, 
          in particular the diagonal and anti-diagonal and the medial point mentioned above, see Fig.~\ref{fig:emp_cop}. 
          It was also noted there that highly correlated stock pairs are ``more elliptical'', 
          and that in periods of high turmoil like the financial crisis, stock pairs are both more correlated and more elliptical, 
          revealing a strong exposure to a common mode of volatility.          
    \item Predict the structure of non-linear (absolute values and quadratic) correlations with a reduced number of parameters, 
          in order to clean the empirically measured dependence coefficients and allow for efficient out-of-sample risk control.
\end{itemize}

As we shall show below, we achieve this with a ``nested'' factor model, i.e.\ standard factor model with volatilities of the factors that 
have themselves a multiplicative factor structure.  We establish that the factor model for factor-volatilities requires 
one (or perhaps two) ``dominant mode'' (that also contributes to the volatility of the residuals) plus idiosyncratic contributions. 
Perhaps surprisingly, this dominant volatility mode is {\it not} the volatility of the dominant (market) mode of the linear factor model.

Several very recent studies have reached conclusions that partly overlap  ours. 
\cite{kelly2012volatility_art} document strong comovements of individual stock return volatilities.
They find that the residuals of factor models (like \cite{fama1993common} or Principal Components Analysis) exhibit 
a strong volatility dependence, which they capture using a one-factor (vol) model.
This is to our knowledge the only attempt in the equity literature at describing volatility dependences in stock returns 
as a second-order effect, after removing linear correlations. 
However, it only focuses on \emph{residuals} volatility, and thus misses the \emph{factors} volatility correlations, as we reveal below.
As mentioned when discussing Fig.~\ref{fig:payoff_option}, the options community is also much concerned by
volatility dependences for the risk description of options portfolios, and some studies have begun to address this issue.
Notably, \cite{engle2012modeling_art} acknowledge the comovements of the implied volatilities of options on individual stocks, 
and attempt to model their dynamics through an exposure to the VIX index.
Very recently, \cite{christoffersen2013factor_art} have proposed to calibrate a one-factor model with option data, 
using its predictions in terms of option pricing.
They model stock prices dynamics with an exposure to a common stochastic market factor 
and stochastic idiosyncratic volatilities with correlated innovations.
Our scope is rather multivariate analysis and bottom-up copula description. 
Our model focuses solely on cross-sectional properties and has for now no dynamical content, 
although this is a natural next step, which is easy to do (at least conceptually).
Our nested factor is more complete than the ones mentioned above, in that it is able to reproduce more stylized facts, 
in particular the common structure of factors volatilities \emph{and} idiosyncratic volatilities, and the 
subtle properties of the bivariate empirical copulas.

\subsubsection*{Data set}
We will construct and calibrate our model on the daily close-to-close log-returns of stock prices of companies 
that are present in the S\&P500 index during the whole of the period studied.
We will be considering three periods of roughly 5 years for the empirical study and the model calibration: 
before the financial crisis (Jan, 2000 -- Dec, 2004);
during the financial crisis (Jan, 2005 -- Dec, 2009);
after  the financial crisis (Aug, 2009 -- Dec, 2012).
A longer dataset is used for the sliding windows procedure of In-sample/Out-of-sample testing 
in the last section: there we consider the ten years period 2000--2009.

It will be useful to group the companies according to their sector of activity,
in order to see if decipherable patterns appear. 
We will make use for that purpose of Bloomberg's classification as summarized in Tab.~\ref{tab:sectors}.
\begin{table}
    \center
    \caption{Economic sectors according to Bloomberg classification, 
             with corresponding number of individuals for each period.}
    \begin{tabular}{lc||c|c|c||c|}
        Bloomberg sector            & Code          & 2000--04 & 2005--09& 2009--12 & 2000--09\\\hline\hline
       %Asset Backed Securities     & \# 1          &   0    &   0     &          &   0\\
       %Basic Materials             & \# 2          &   0    &   0     &          &   0\\
        Communications              & \# 3          &  33    &  25     &  29      &  18\\
        Consumer, Cyclical          & \# 4          &  60    &  49     &  33      &  40\\
        Consumer, Non-Cyclical      & \# 5          &  67    &  75     &  75      &  53\\
       %Diversified                 & \# 6          &   0    &   0     &   1      &   0\\
        Energy                      & \# 7          &  19    &  21     &  34      &  15\\
        Financial                   & \# 8          &  57    &  55     &  75      &  37\\
       %Funds                       & \# 9          &   0    &   0     &   0      &   0\\
       %Government                  & \#10          &   0    &   0     &   0      &   0\\
        Industrial                  & \#11          &  51    &  50     &  50      &  42\\
       %Mortgage Securities         & \#12          &   0    &   0     &   0      &   0\\
        Technology                  & \#13          &  38    &  43     &  35      &  33\\
        Utilities                   & \#14          &  27    &  27     &  28      &  24\\\hline
        Total number of firms ($N$) &               & 352    & 345     & 359      & 262\\\hline
        Total number of days ($T$)  &               &1255    &1258     & 755      &2514
    \end{tabular}
    \label{tab:sectors}
\end{table}

\subsubsection*{Outline}

This paper is made of an introduction, four sections and a conclusion. 
In Section~\ref{sec:MFlin}, we study the linear correlations of pairs of stocks and 
discuss the design and estimation of a factor model for their description.
The non-linear dependences of factors and residuals generated by the calibration of the model are
studied in Section~\ref{sec:MFspectral}, and motivate  the specification of the volatility content
of the model that we present in Section~\ref{sec:modeling_vol}.
The resulting nested factor, non-Gaussian model is calibrated, and Section~\ref{sec:stability} is dedicated to an Out-of-Sample stability analysis,
that validates the usefulness of our description of non-linear dependences.
Methodological points for the estimation of the model parameters and the assessment of the model's performance are provided in appendices.

\section{Linear factors}\label{sec:MFlin}

We first recall the definition and basic properties of a simple one-level factor model for the joint description of the stock returns $\ret_i$ of $N$ firms, 
as a combination of $M$ shared factors $f_k$:
\begin{equation}\label{eq:MODEL}
    \ret_i=\sum_{k=1}^M \Wei_{ki}f_k+e_i.
\end{equation}
The weight $\Wei_{ki}$ parameterizes the linear exposure of stock $i$ to factor $k$.
At this stage, we do not yet specify the statistical properties of the factors $f_k$ and residuals $e_i$,
except that we impose that they are {\it linearly uncorrelated}, i.e.:
\begin{subequations}\label{eq:ortho_f_e}
\begin{align}
    \esp{f_kf_\ell}&=\delta_{k\ell}\\
    \esp{e_ie_j}&=\delta_{ij}\,\Big(1-\sum_\ell\Wei_{\ell i}^2\Big)\\
    \esp{f_ke_j}&=0.
\end{align}
\end{subequations}
Written in matrix form, the factor model reads:
 \begin{equation}
 \label{eq:MODEL_mat}\tag{\ref{eq:MODEL}$'$}
 \Ret=\mat{F}\Wei+\mat{E},
 \end{equation}
with unit-variance factors $\mat{F}$ $(T\times M)$, exposures $\Wei$ $(M\times N)$ of every stock to every factor, 
and orthogonal residuals $\mat{E}$ $(T\times N)$.
In this way, the residuals $e_i$ can be understood as idiosyncratic shocks and all the linear dependence is 
accounted for by the factors. The predictions of the model in terms of covariances of the returns $\ret_i$ do not need additional assumptions,
and only depend on the matrix of linear weights $\Wei$. Assuming that the returns are normalized to have unit variance, one has: 
\begin{equation}\label{eq:lincor_predict}
    \rho_{ij}=\esp{\ret_i \ret_j}=\begin{cases}(\Wei^{\dagger}\Wei)_{ij}&,\, i\neq j\\1&,\, i=j\end{cases}
\end{equation}

The above linear factor model is of course the workhorse of the econometric literature. However, there are two subtle points about it that need 
to be clarified.
\begin{enumerate}

\item In the econometric literature, one often assumes that the set of explanatory factors is known. The time series 
of these factors $\mat{F}_{tk}$ are then {\it inputs} of the estimation problem, whereas the elasticities $\Wei$ are the output 
of the linear regression. Here, we will rather determine the weights $\Wei$ in such a way that the empirical correlation 
matrix is as close as possible to the one predicted by an $M$-factor model, Eq.~\eqref{eq:lincor_predict}:
\begin{equation}\label{eq:offdiag_content}
    \argmin \left\|\frac{1}{T}\Ret^\dagger \Ret-\Wei_{\text{F}}^\dagger \Wei_{\text{F}}\right\|_{\text{off-diag}}.
\end{equation}
When the weights $\Wei_{\text{F}}$ are known, it is possible to design a different identification scheme
that generates orthogonal residuals. Consider indeed the date-by-date regression of the $N$ returns 
on the $M$ (freshly estimated) weights $\Wei_{\text{F}}$:
\begin{equation}\label{eq:dbdR}
    \Ret_{t\cdot}=\mat{F}_{t\cdot}\Wei_{\text{F}}+\mat{E}_{t\cdot}.
\end{equation}
The regression parameters to be estimated are then the value of the $M$ factors $\mat{F}_{t\cdot}$ for date $t$.
A GLS solution of the regression then yields the wanted factors and residual series.
It is only {approximate} in the sense that $\frac{1}{T}\mat{E}^\dagger \mat{E}$ is only ``as close as can be'' to a diagonal matrix,
and $\frac{1}{T}\mat{F}^\dagger \mat{F}$ is only approximately $\mathds{1}_M$.

\item The linear factor methodology looks superficially similar to a standard Principal Components Analysis (PCA).
We expand in Appendix~\ref{app:PCA} on the similarities and differences between the two points of view. In fact, we use the results of the PCA as a 
starting point for the numerical 
optimization program defined by Eq.~\eqref{eq:offdiag_content}. 

\end{enumerate}

We will show in  Sect.~\ref{sec:stability} that our factor model approach in fact outperforms the PCA approach 
by more than 5\% when it comes to comparing the out-of-sample risk of optimal portfolios constructed using these  
two methods\footnote{The PCA method is also known as eigenvalue clipping in the context of cleaning schemes for matrix inversion,
 and is one of the best generic cleaning scheme known so far, see \cite{potters2009financial}.}, 
with an in-sample risk almost unchanged, see Fig.~\ref{fig:ISOS_lin} below.

We have calibrated the linear weights $\Wei_{\text{F}}$ on the three data sets by solving Eq.~\eqref{eq:offdiag_content}, 
using $M=10$. We will discuss in details in the next section the properties of the factors time series $\mat{F}_{t\cdot}$ 
and residual time series $\mat{E}_{t\cdot}$; we will show in particular that while these time series are indeed approximately 
uncorrelated, strong non-linear dependencies remain, and this will suggest the building blocks of our nested factor model.
%%% CODE R:
%%% cor(estimQ2mf[["F"]])[lower.tri(diag(M))]
%%% cor(estimQ2mf[["E"]])[lower.tri(diag(S))]

\section{Properties of the reconstructed factors and residuals}\label{sec:MFspectral}

The calibration procedure of the linear model~\eqref{eq:MODEL_mat} worked out in the previous section 
outputs the series of factors $\mat{F}_{t\cdot}$ and residuals $\mat{E}_{t\cdot}$.
The average linear correlation over all pairs of factors is indeed very small, $\approx 1.2\cdot 10^{-4}$, with a standard-deviation $55\cdot10^{-4}$, 
similar for all periods. The average linear correlation over all pairs of residuals is around $(-2.5\pm 45)\,10^{-3}$, and
the average cross-correlation between factors and residuals is of the order $10^{-4}$.
Although not exactly zero these small numbers are clearly within the noise 
(which is larger than $1/\sqrt{T}\approx 3\cdot 10^{-2}$ because of volatility persistence) 
and illustrate that the resulting series of factor returns and residuals are 
to a very good approximation all \emph{uncorrelated}.

This does not mean however that they are \emph{independent}. Indeed, we will show in this section that all the volatilities of these series are 
strongly dependent. We will therefore enhance the factor model defined in Eqs.~(\ref{eq:MODEL},\ref{eq:ortho_f_e}) by a characterization 
of the non-linear dependences among the $f$'s and the $e$'s. Note that the factors and residuals are expected (and found) to be strongly
non-Gaussian. In fact, we will model the volatilities as approximately log-normal processes (but see below).

The non-linear properties of the reconstructed factors and residuals can be investigated through the correlations of 
absolute values, or squares, etc. Since this choice is to some extent arbitrary, we have defined the generalized 
non-linear correlations for factors and residuals as:
\begin{subequations}\label{eq:allcorabs}
\begin{align}
    \label{eq:facfac}C^{\ff}_{k\ell}(p)=\frac{1}{p^2}\ln\frac{\vev{|\mat{F}_{tk}\mat{F}_{t\ell}|^p}}{\vev{|\mat{F}_{tk}|^p}\vev{|\mat{F}_{t\ell}|^p}}\\
    \label{eq:resres}C^{\rr}_{ij}   (p)=\frac{1}{p^2}\ln\frac{\vev{|\mat{E}_{ti}\mat{E}_{tj}|^p}}{\vev{|\mat{E}_{ti}|^p}\vev{|\mat{E}_{tj}|^p}}\\
    \label{eq:facres}C^{\fr}_{kj}   (p)=\frac{1}{p^2}\ln\frac{\vev{|\mat{F}_{tk}\mat{E}_{tj}|^p}}{\vev{|\mat{F}_{tk}|^p}\vev{|\mat{E}_{tj}|^p}}
\end{align}
\end{subequations}
for any value of $p>0$. The normalization with $p^{-2}$ ensures that these coefficients tend to a finite value when $p \to 0$, and
they would actually be independent of $p$ for multivariate log-normal volatilities. Large values of $p$ lead to very noisy estimators, so 
we restrict below to $p \in (0,2]$.

\begin{figure}
    \center
    \includegraphics[scale=.55,trim=  0 165  710 0,clip]{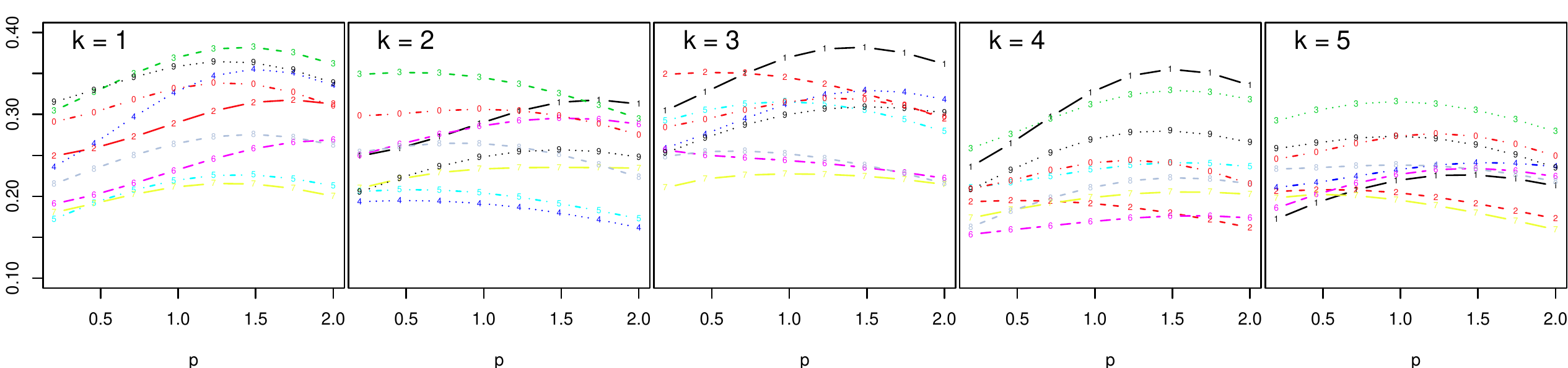}
    \includegraphics[scale=.55,trim=  0 165 1416 0,clip]{imagesPDF/2000-2009_emp_logcorabs_fact.pdf}%
    \includegraphics[scale=.55,trim=733 165    0 0,clip]{imagesPDF/2000-2009_emp_logcorabs_fact.pdf}
    \caption{Visual representation of the estimated factors-factors dependences, for $M=10$, on the period 2000--2009. 
    The correlation~\eqref{eq:facfac} is shown for every factor $k$ with all other factors $\ell \neq k$,
    as a function of the order $p$ of the absolute moment considered.}
    \label{fig:matelems}
\end{figure}

As an example, we show in Fig.~\ref{fig:matelems} the off-diagonal matrix elements of the 
factor-factor correlations \eqref{eq:facfac} for $M=10$. Each figure corresponds to a value of $k$, the different 
curves represent the values of Eq.~\eqref{eq:facfac} for different $\ell \neq k$, as a function of $p$.
We observe that a) these correlations clearly are non-zero, whereas a factor model with Gaussian statistics would 
give zero (since in this case, factors would not only be uncorrelated but independent); 
and b) The concavity of the curves is a signature of 
non-Gaussianity in log-volatilities, while their splitting (in particular as $p\to 0$) reveals a complex structure 
that we will uncover using a model in Section~\ref{sec:modeling_vol} below.

Because of the large number of residuals, the above ``naked eye'' analysis is not possible for the 
factors-residuals and residuals-residuals correlations, for which it turns much more convenient to use a spectral approach
in terms of singular value decompositions (which boils down to eigenvalues/eigenvectors for symmetric objects like $C^{\ff}$ and 
$C^{\rr}$). In terms of singular values/eigenvalues, we find that for all $p \in (0,2]$, two of them clearly stand out, while the rest 
stands within a noisy ``bulk''.\footnote{A third eigenvalue of $C^{\rr}$ might in fact be significant, but we will discard it altogether in the present 
study.} The largest one is furthermore a factor 3 to 5 larger than the second one, suggesting that a one- (or two-) factor
model for the log-volatilities should provide a good description of the data (see below). 

\begin{figure}
    \center
 
    \label{fig:evects} 
    \includegraphics[scale=.75,trim=  0 138  600   0,clip]{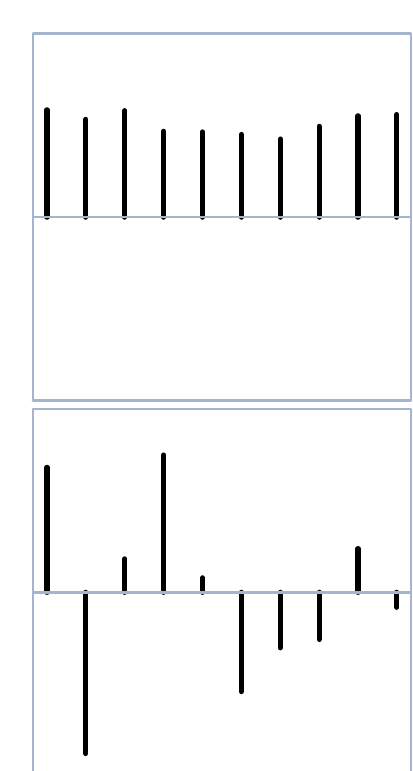}\hspace{.8cm}
    \includegraphics[scale=.75,trim=120 138  480   0,clip]{imagesPDF/2000-2009_svd_vects.pdf}\\
    \includegraphics[scale=.75,trim=  0   0  600 331,clip]{imagesPDF/2000-2009_svd_vects.pdf}\hspace{.8cm}
    \includegraphics[scale=.75,trim=120   0  480 331,clip]{imagesPDF/2000-2009_svd_vects.pdf}
    \caption{$M=10$, 2000--2009. First two eigenvectors, from top to bottom, of the 
        factor-factor (right), and 
    residual-residual correlations (left, stock indices ordered in different Bloomberg sectors, separated by vertical grey lines).}
\end{figure}

The corresponding eigenvectors are to a good approximation independent of $p$. 
The components of two dominant eigenvectors of $C^{\ff}$ are shown in Fig.~\ref{fig:evects}(left), for the period 2000--2009 and averaged over $p$. 
As expected, the largest eigenvalue has an associated vector approximately uniform over the $M=10$ factors. 
Zooming into sub-periods, this mode seems to be impacting/impacted by the financial sector more strongly 
in the 2005--2009 period (containing the financial crisis) whereas in other periods it is almost uniformly spread over sectors.
The second eigenvalue has a non-trivial structure which is less robust in details,
although the overall pattern is similar for the different sub-periods.

The eigenvectors of $C^{\rr}$ are of dimension $N$ (the number of stocks) and thus 
less easy to visualize. We show in Fig.~\ref{fig:evects}(right) the components of its two dominant eigenvectors in a representation where stocks are grouped 
according to their Bloomberg classification (see the grey vertical lines separating these sectors). Note that the finance sector plays a special role here:
its weight is larger in the first eigenvector, while the second eigenvector is to a first approximation ``finance against all''. Zooming again into sub-periods,
the financial sector is clearly a stand-alone mode of fluctuations in the crisis period. In the pre-crisis period, the second relevant mode is rather composed of commodities. 
Indeed, the second eigenvector features the opposition of utilities, energy and communications against the rest. In the post-crisis period, on the 
other hand, there is no clear signature of the structure of the second eigenvector of $C^{\rr}$.

Finally, the singular value  decomposition of the mixed correlation matrix $C^{\fr}$ is consistent with the above findings: the two dominant left-eigenvectors 
are nearly identical 
to the two dominant eigenvectors of $C^{\ff}$ while the two dominant right-eigenvectors are nearly identical to the two dominant eigenvectors of $C^{\rr}$. 
This confirms that we only need to focus on these four eigenvectors, two of dimension $M=10$, two of dimension $N$.

\section{A factor model for volatilities}\label{sec:modeling_vol}

The spectral analysis of the previous section suggests the existence of {\it two volatility factors} that drive the amplitude of {\it both} the factors $f_k$ 
and the residuals $e_i$.

More concretely, we propose the following multiplicative model for the volatilities, which defines our {\it nested factor model}:
\begin{subequations}\label{eq:model_fkej_0}
\begin{align}
    f_k&=\epsilon_k \,\exp(A_{k0} \Omega_0 + A_{k1} \Omega_1 +  \omega_k)\\\label{eq:model_ej_0}
    e_j&=\eta_j     \,\exp(B_{j0} \Omega_0 + B_{j1} \Omega_{1} +  \widetilde\omega_j),
\end{align}
\end{subequations}
where the $\Omega$'s are stochastic {\it factor} log-volatilities and the $\omega$'s are stochastic ``idiosyncratic'' log-volatilities 
(all independent of each other and independent of the Gaussian noises $\epsilon$'s and $\eta$'s). 
The parameters $A$'s and $B$'s weight the contribution of every volatility mode.
In particular, we expect $A_{k0}$ and $B_{j0}$ to be given by the dominant eigenvectors of $C^{\ff}$ and $C^{\rr}$, respectively, 
and $A_{k1}$ and $B_{j1}$ by the second eigenvectors. 

In the next subsection, we first estimate a minimal model with a single volatility driver, $\Omega_0$.

\subsection{A dominant volatility mode}

\subsubsection{Definition}

The minimal improvement over the independent factors assumption, while keeping {\it uncorrelated} factors,
is to allow for a single common source for the fluctuation of amplitudes, i.e.\ set $A_{k1}=B_{j1}=0$, $\forall k,j$ in the above equations:
\begin{subequations}\label{eq:model0}
\begin{align}
    f_k&=\epsilon_k \,\exp(A_{k0} \Omega_0  +  \omega_k)\\\label{eq:model_ej_0}
    e_j&=\eta_j     \,\exp(B_{j0} \Omega_0  +  \widetilde\omega_j),
\end{align}
\end{subequations}
with $\epsilon_k,\eta_j$ Gaussian, with variance such that Eqs.~\eqref{eq:ortho_f_e} hold. 

Because this model is already a level of complexity higher than the standard linear factor model, it is worthwhile to 
insist on the intuitive meaning of the different log-volatility factors:
\begin{align*}
    \Omega_0 &= \text{dominant and common driver of log-volatility across all factors and residuals,} \\
    \omega_1 &= \text{idiosyncratic log-volatility of  the market mode $f_1$ (a.k.a.\ the index), net of $\Omega_0$,}
\end{align*}
and the subsequent $\omega_k, \widetilde\omega_j $ ($k>1$) characterize the ``residual volatilities''
not explained by the common driver $\Omega_0$ in the amplitude of the factors $f_k$ and residuals $e_j$. 
Note in particular that the dominant log-volatility factor $\Omega_0$ cannot be identified with the log-volatility
of the dominant market mode $f_1$ in the linear factor model! 

The model is completely characterized from a probabilistic point of view 
when the law of the log-volatilities is specified.
The $p$-dependence of the curves in Fig.~\ref{fig:matelems} suggests that the non-Gaussianity in the log-volatilities 
is approximately homogeneous across the factors, and thus possibly due to the common volatility driver $\Omega_0$ alone,
while the residual volatilities $\omega_k$ and $\widetilde\omega_j$ can be taken as Gaussian (at least in a first approximation).
For $\Omega_0$, we set:
    \[
    \esp{\Omega_0  }=0\qquad
    \esp{\Omega_0^2}=1\qquad
    \esp{\Omega_0^3}=\zeta_0\qquad
    \esp{\Omega_0^4}=3+\kappa_0.
    \]

At this stage, a recap is probably useful. Our nested factor model with a single volatility mode (defined by Eqs.~(\ref{eq:MODEL},\ref{eq:model0})) 
contains the following parameters:
\begin{itemize}
\item  $MN$ linear weights $\Wei_{ki}$ (already estimated, see Sect.~\ref{sec:MFlin});
\item $M$ coefficients $A_{k0}$ and $N$ coefficients $B_{j0}$ giving the exposure of factors and residuals to the common volatility mode $\Omega_0$ (of unit variance);
\item The standard-deviations $s_k$ and $\widetilde s_j$ of the residual Gaussian log-volatilities $\omega_k$ and $\widetilde\omega_j$;
\item And finally the skewness $\zeta_0$ and kurtosis $\kappa_0$ of the dominant volatility mode $\Omega_0$.
\end{itemize}
    So there are overall $NM+2(N+M)+2$ parameters, for a dataset of size $NT$. 
    More importantly the number of parameters is only marginally increased with respect to a typical linear factor model 
    (where only the $NM$ linear weights enter into account): only $2(N+M+1)$ new parameters, intended to 
    improve the description of {\it all} $N(N-1)/2$ pairwise dependences coefficients.
    
The calibration procedure, that allows to determine these $2(N+M+1)$ new parameters, is detailed in Appendix~\ref{app:calibVol}. 

\subsubsection{Results of the calibration}

\begin{figure}
    \center
    \subfigure[$A_{k0}$ and $s_{k}$]{\includegraphics[scale=.45]{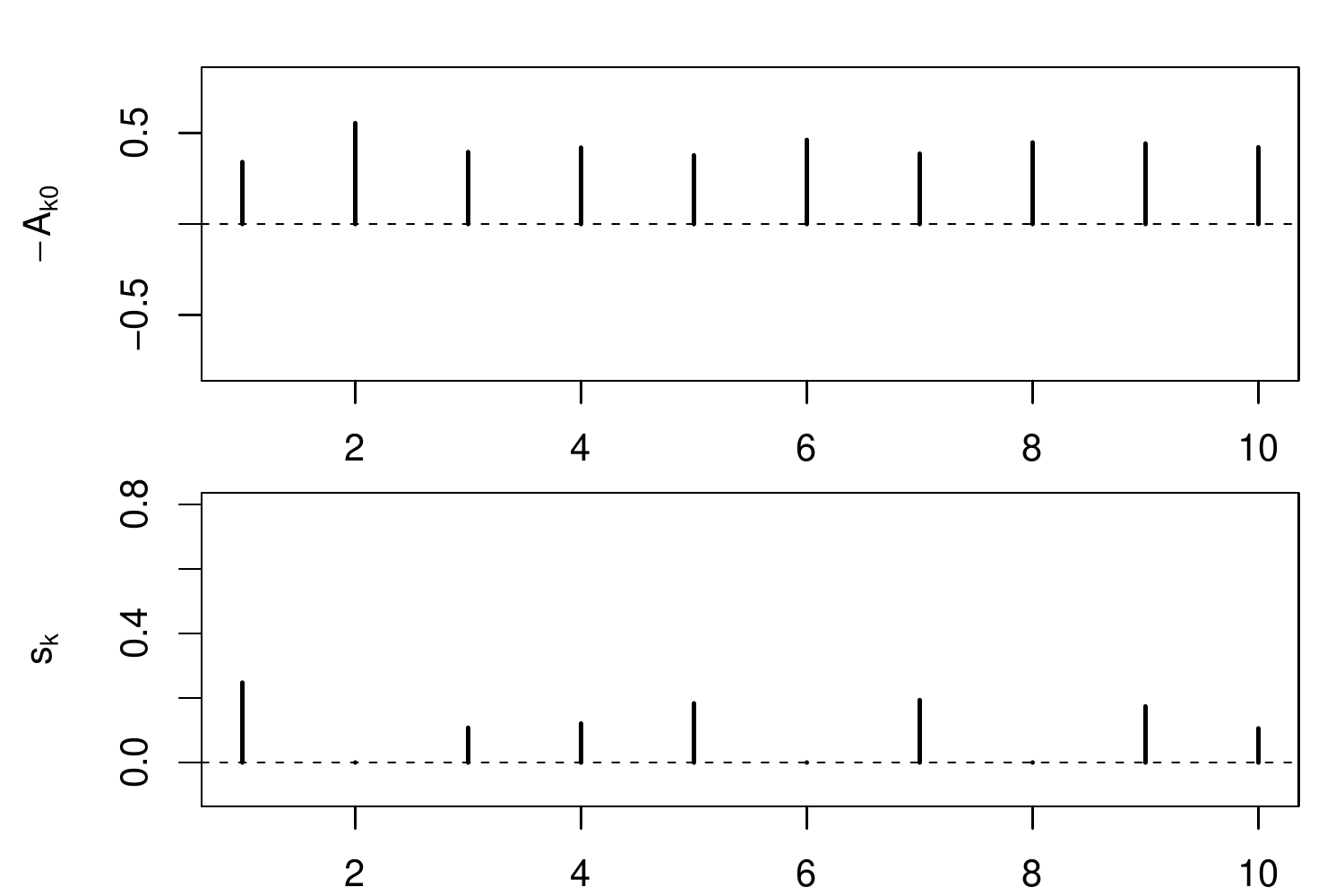}} %\hspace{.5cm}
    \subfigure[$B_{j0}$ and $\widetilde s_{j}$]{\includegraphics[scale=.45]{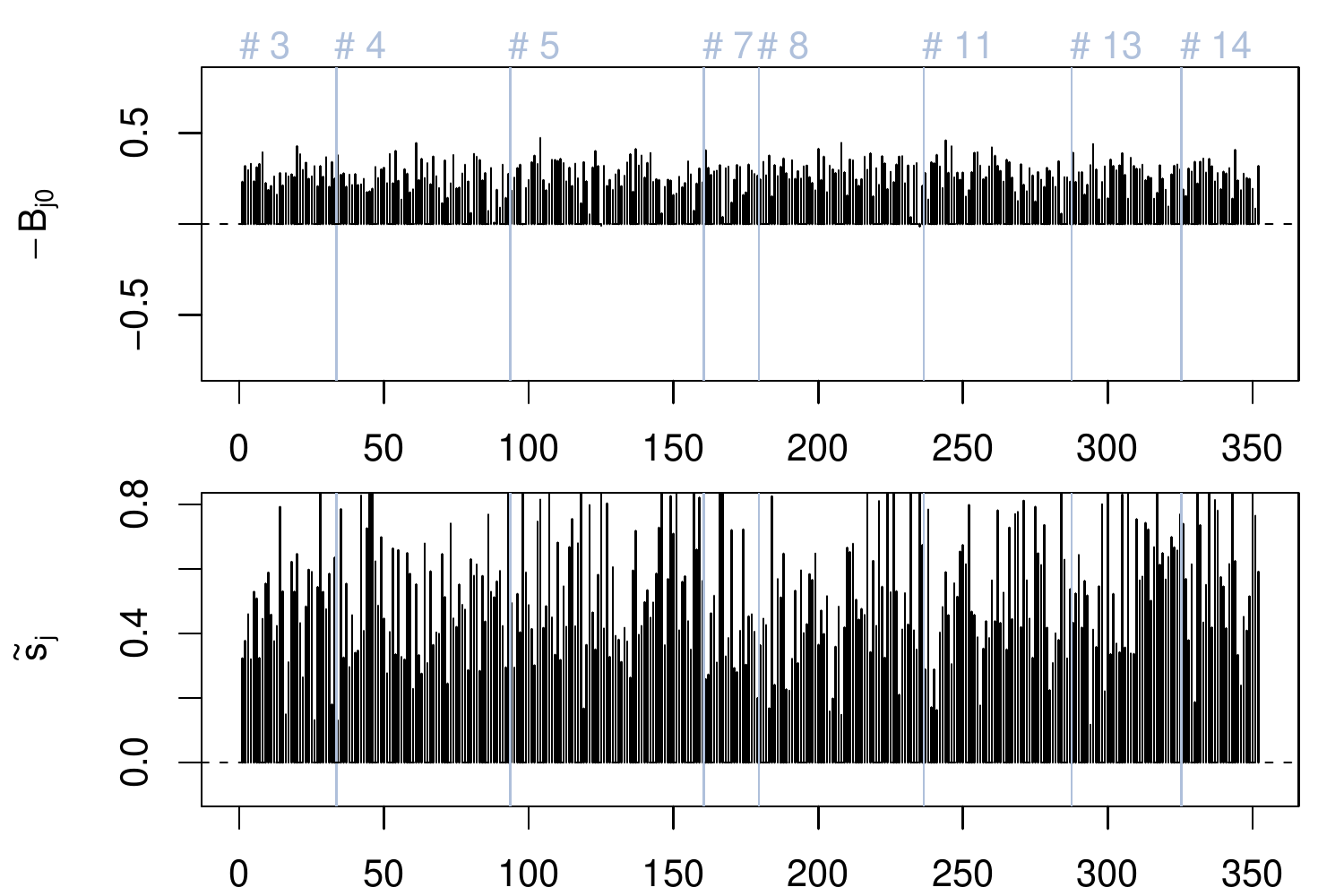}} %\hspace{.5cm}
    \caption{Estimated volatility factor loadings and volatility residuals. $M=10$, 2000--2004.}\label{fig:AB2000-2004}
    \subfigure[$A_{k0}$ and $s_{k}$]{\includegraphics[scale=.45]{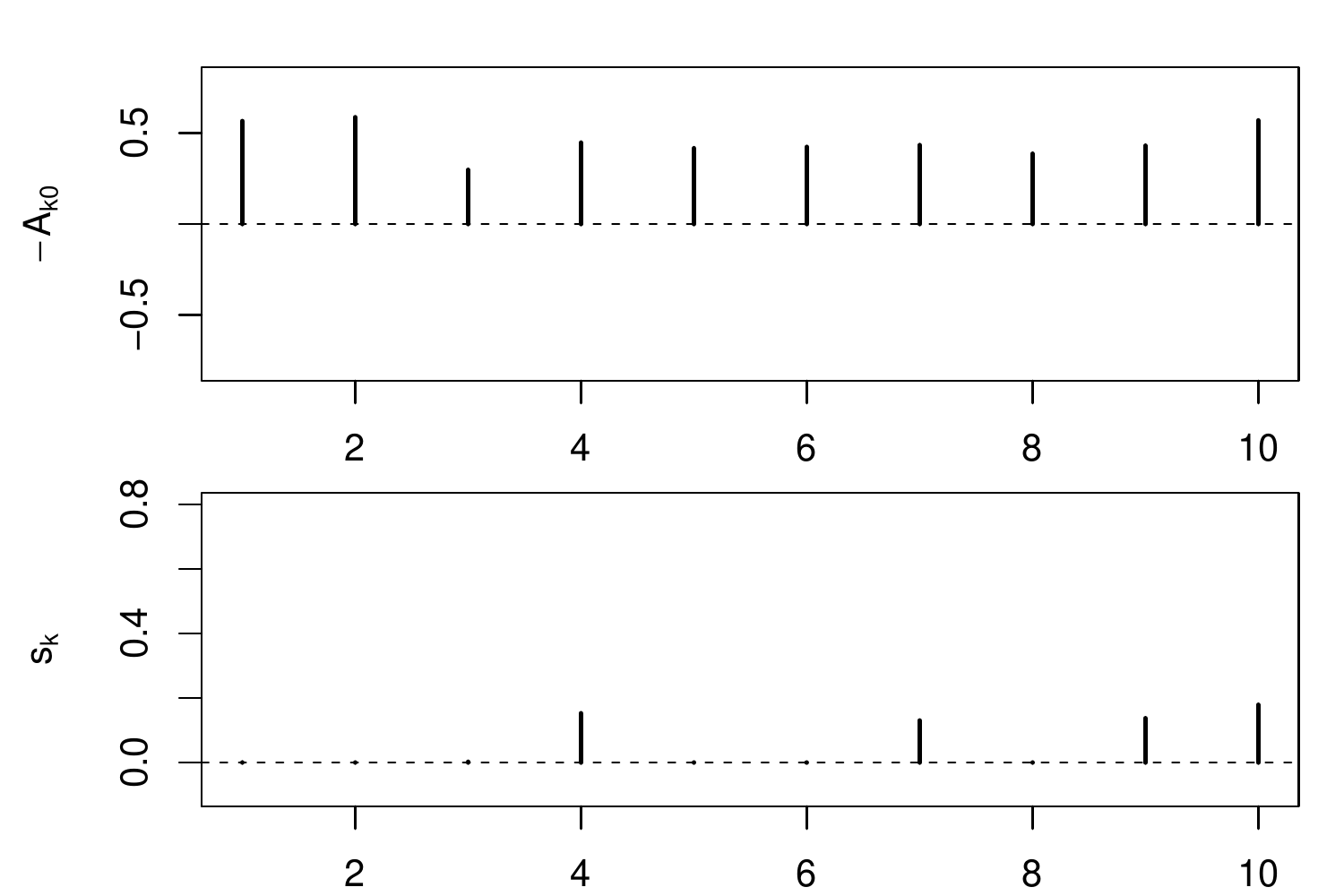}} %\hspace{.5cm}
    \subfigure[$B_{j0}$ and $\widetilde s_{j}$]{\includegraphics[scale=.45]{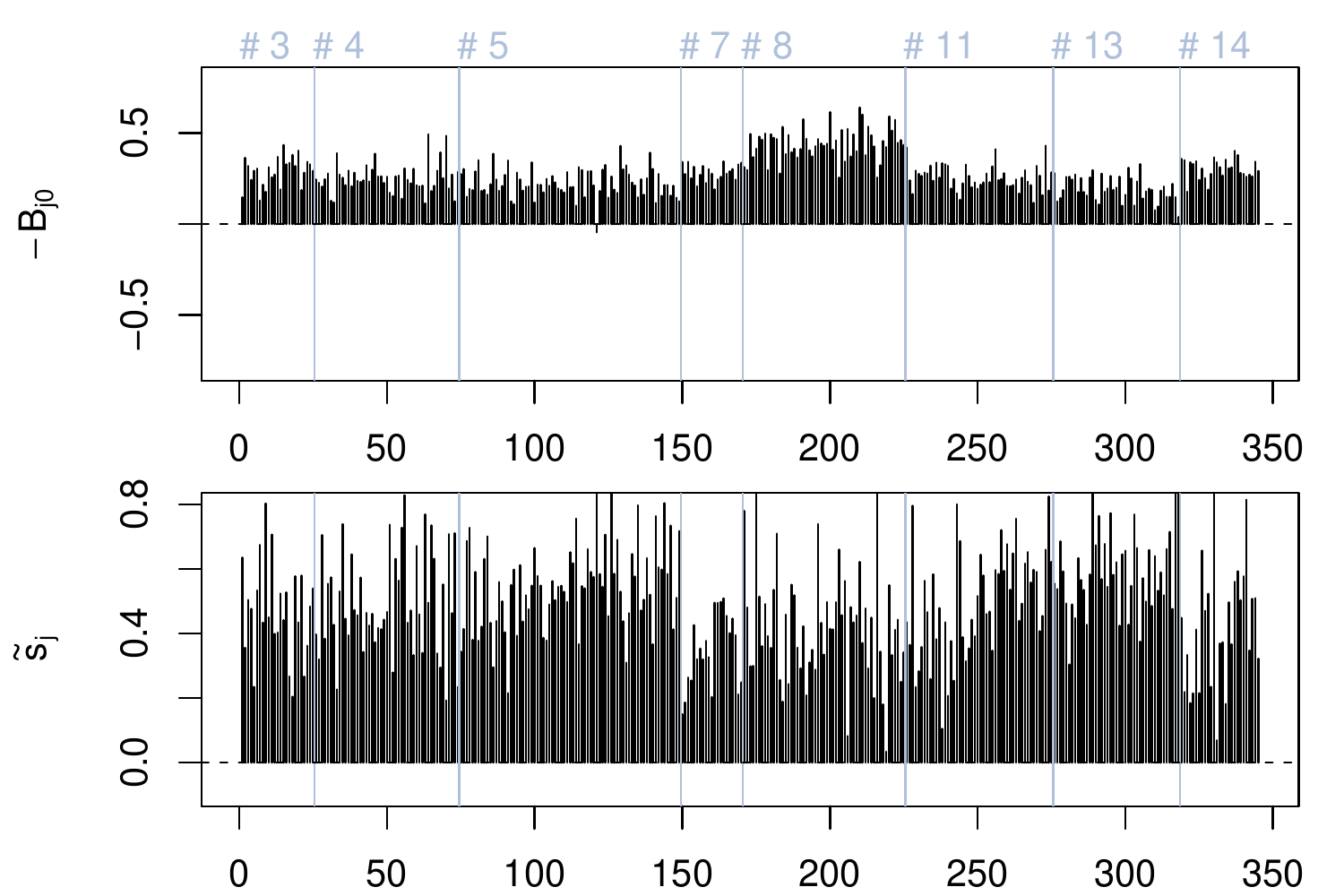}} %\hspace{.5cm}
    \caption{Estimated volatility factor loadings and volatility residuals. $M=10$, 2005--2009}\label{fig:AB2005-2009}
    \subfigure[$A_{k0}$ and $s_{k}$]{\includegraphics[scale=.45]{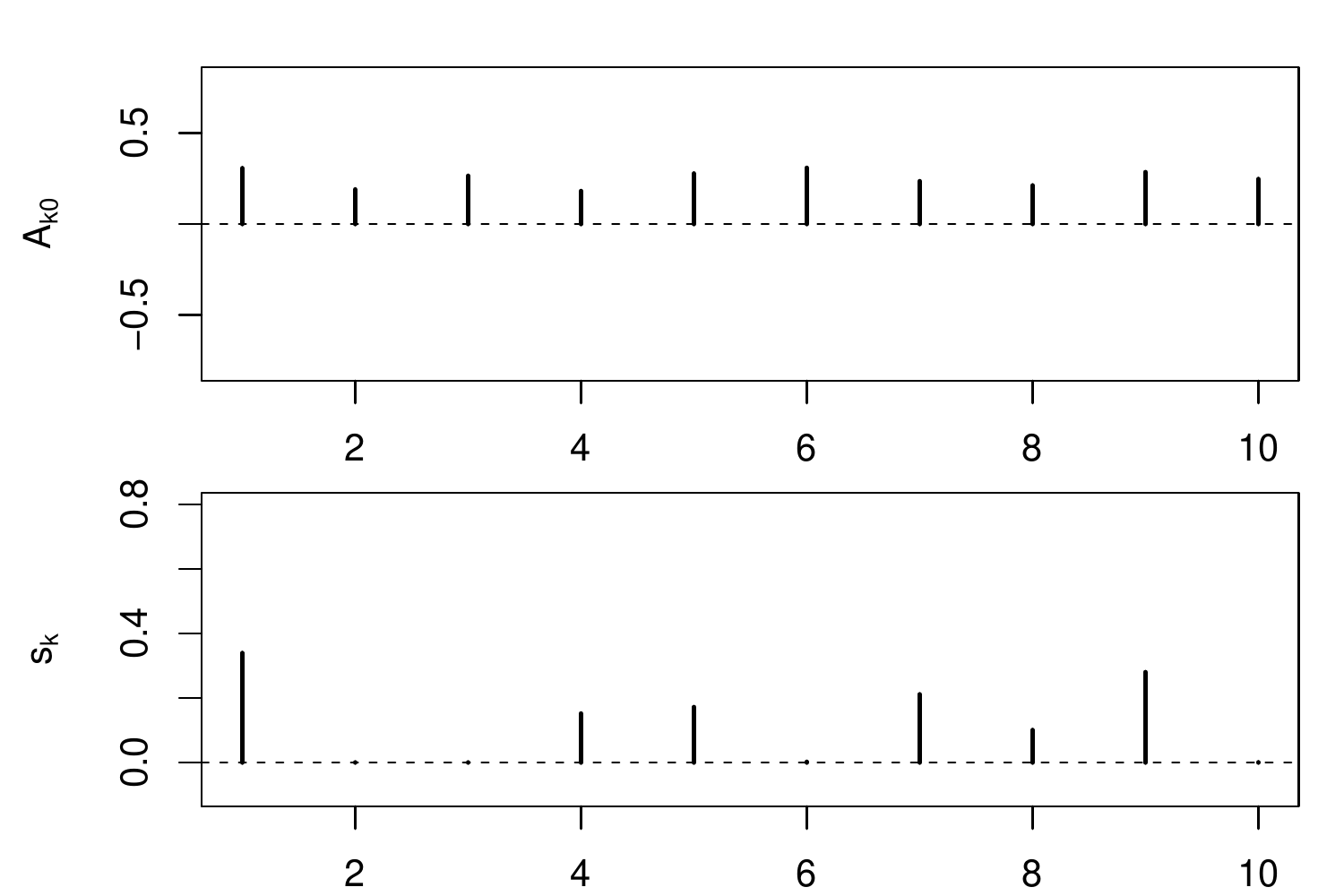}} %\hspace{.5cm}
    \subfigure[$B_{j0}$ and $\widetilde s_{j}$]{\includegraphics[scale=.45]{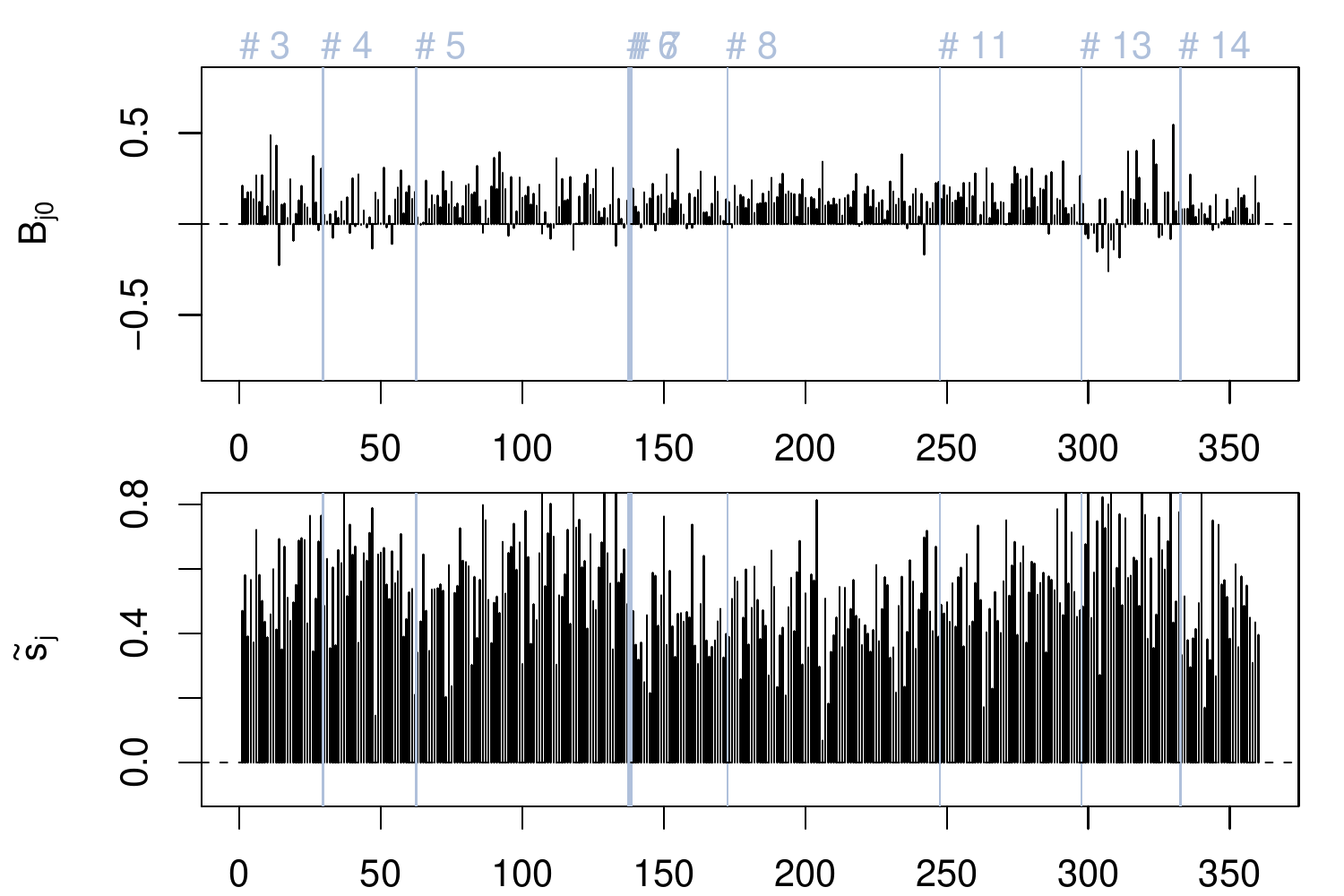}} %\hspace{.5cm}
    \caption{Estimated volatility factor loadings and volatility residuals. $M=10$, 2009--2012}\label{fig:AB2009-2012}
\end{figure}

The calibration results are given graphically in Figs.~\ref{fig:AB2000-2004}, \ref{fig:AB2005-2009}, \ref{fig:AB2009-2012},
    where we show, separately for each sub-period, the estimated parameters $A_{k0}$ and $B_{j0}$.
    For the reason discussed in Appendix~\ref{app:calibVol}, they turn out to be very close to the first eigenvector of 
    the corresponding matrix of ``log-abs'' correlations discussed in Sect.~\ref{sec:MFspectral} above.
    Of particular interest are the ratios $B_{j0}/A_{10}$, which are found to be on average less than unity (0.79 in 2000--2004, 0.49 in 2005--2009 and 0.40 in 2009--2012). 
    This means that the dominant volatility mode affects both the index volatility and the residual volatilities, as noted in \cite{cizeau2001correlation}, 
    but in a  weaker way for the latter. This was already observed in \cite{allez2011individual}, see their Fig.~3. 
    
    We also show in Figs.~\ref{fig:AB2000-2004}, \ref{fig:AB2005-2009}, \ref{fig:AB2009-2012} the parameters $s_{k}$ and $\widetilde s_{j}$. 
    Note that some factors $k$ seem to have their volatility {\it entirely explained} by the common driver $\Omega_0$
    so that there is no residual volatility left.
    The estimated values of the non-Gaussianity parameters of the log-volatilities are reported in Tab.~\ref{tab:ILE}.
    Noticeably, the kurtosis of the common driver $\Omega_0$ is found to be {\it negative} in every period:
    the log-volatility is less kurtic than a Gaussian, which is a rare finding in financial time series analysis! 
    This was already revealed by the concavity of the curves in Fig.~\ref{fig:matelems}.

\begin{table}
    \center
    \caption{$M=10$. Estimated non-Gaussianity parameters of the dominant log-volatility $\Omega_0$.}
    \begin{tabular}{c||r|r|r|}
                                       &2000--2004&2005--2009&2009--2012\\\hline\hline
                             $\zeta_0$ &$-0.072  $&$-1.492  $&$ 0.607$  \\
                            $\kappa_0$ &$-0.129  $&$-1.916  $&$-0.608$  \\\hline
    \end{tabular}
    \label{tab:ILE}
\end{table}

\subsubsection{Dynamics of the common volatility mode}

\begin{figure}
    \center
    \subfigure[2000--2004]{\includegraphics[scale=.62,trim=230 0  200 0,clip]{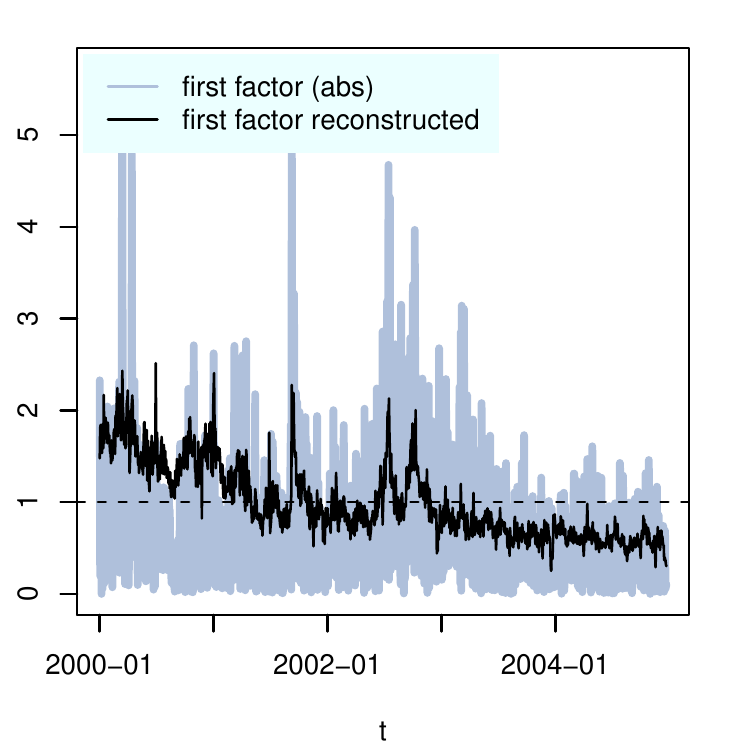}\label{fig:firstfact1mode}}
    \subfigure[2005--2009]{\includegraphics[scale=.62,trim=230 0  200 0,clip]{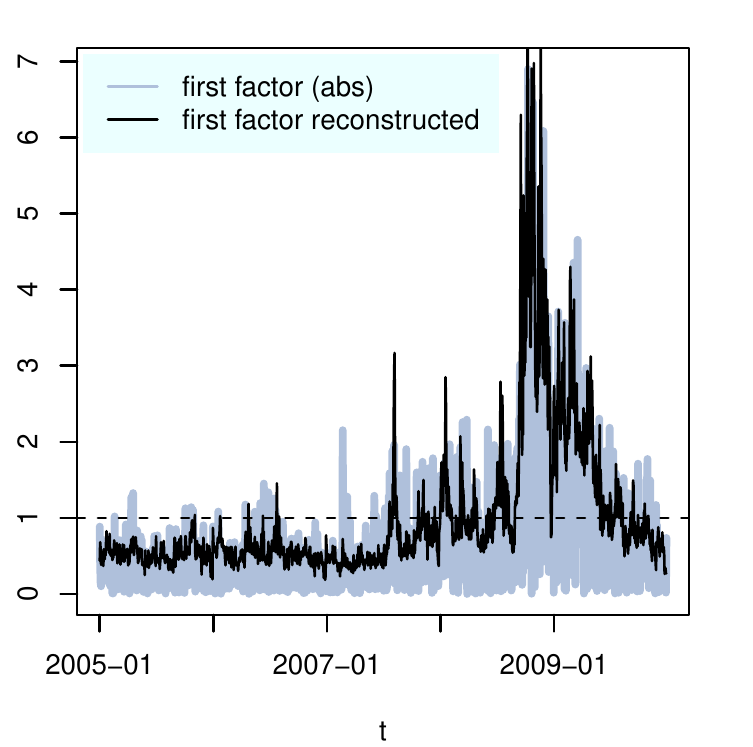}}
    \subfigure[2009--2012]{\includegraphics[scale=.62,trim=230 0  200 0,clip]{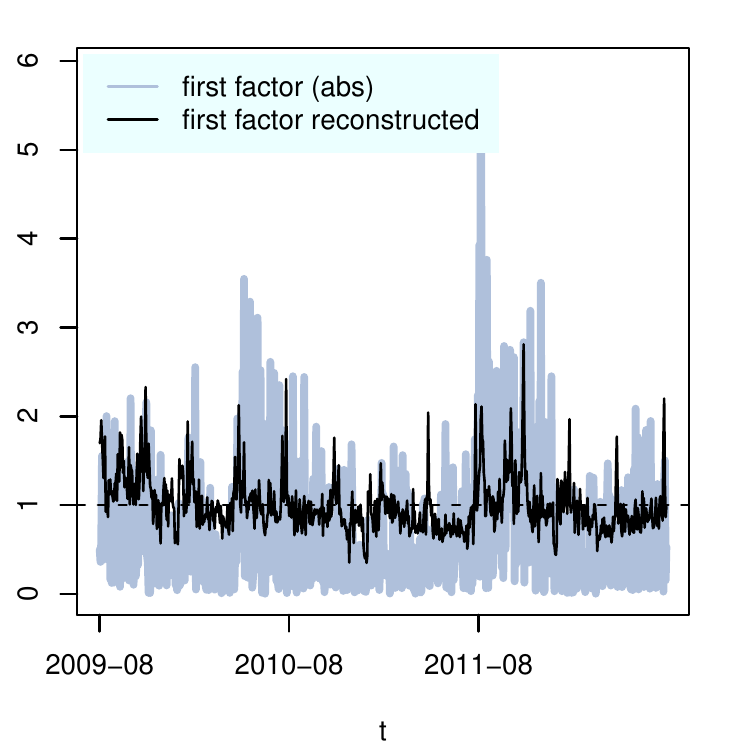}}
    \caption{One volatility driver $\Omega_0$. Original and reconstructed dynamics of the amplitude of the first factor $|f_1|$.
    Note that a second volatility factor $\Omega_1$ and residual volatility $\omega_1$ are needed to improve the match, see below.} 
    \label{fig:firstfact}
\end{figure}

Interestingly, we are now in position to reconstruct the time series of the common volatility mode $\Omega_{t0}$
out of the model equations and the estimated parameters. 
Similarly to what was done in Sect.~\ref{sec:MFlin} to recover the series of linear factors, 
we perform here two date-by-date regressions motivated by the Eqs.~\eqref{eq:model0}:   
\begin{align*}
    \ln|\mat{F}_{tk}|-\vev{\ln|\mat{F}_{tk}|}&=\Omega_{t0} A_{k0} + s_{k} \omega_{tk}\\
    \ln|\mat{E}_{tj}|-\vev{\ln|\mat{E}_{tj}|}&=\Omega_{t0} B_{j0} + \widetilde s_{j} \widetilde\omega_{tj}
\end{align*}
Whereas the first regression is performed over only the $M$ variables $A_{k0}$,
the second one is realized over the $N$ variables $B_{j0}$ and thus leads to much less noisy estimates of $\Omega_{t0}$
(we will always use the second determination in the following).
The overlap of the time series of $\Omega_0$ estimated with the two regressions is nevertheless quite good, 
with a correlation coefficient between 0.55 and 0.75 depending on the period studied.
We show in Fig.~\ref{fig:firstfact} the time series $\exp{(A_{10}\Omega_{t0})}$ 
reconstructed from the procedure above after estimation of the parameters,
that we compare to the absolute value of the market factor $\mat{F}_{t1}$.

An obvious next step would be to calibrate a dynamical model (GARCH or stochastic volatility) to account for the temporal
evolution of $\Omega_{t0}$. 

%\subsection{A dominant volatility mode}

As consistency checks of both the quality of the model and the estimation procedure, we now analyze
the model prediction with the estimated parameters and compare them with empirical measurements of the same quantities.
Of particular interest are the quadratic correlations and the diagonal copulas,
whose anomalies observed in a previous study \citep{chicheportiche2012joint} actually motivated the present model.

\subsubsection{Quadratic correlations}

\begin{figure}[p]
    \center
    \subfigure[2000--2004]{\includegraphics[scale=.2,trim=750 50 0 50,clip]{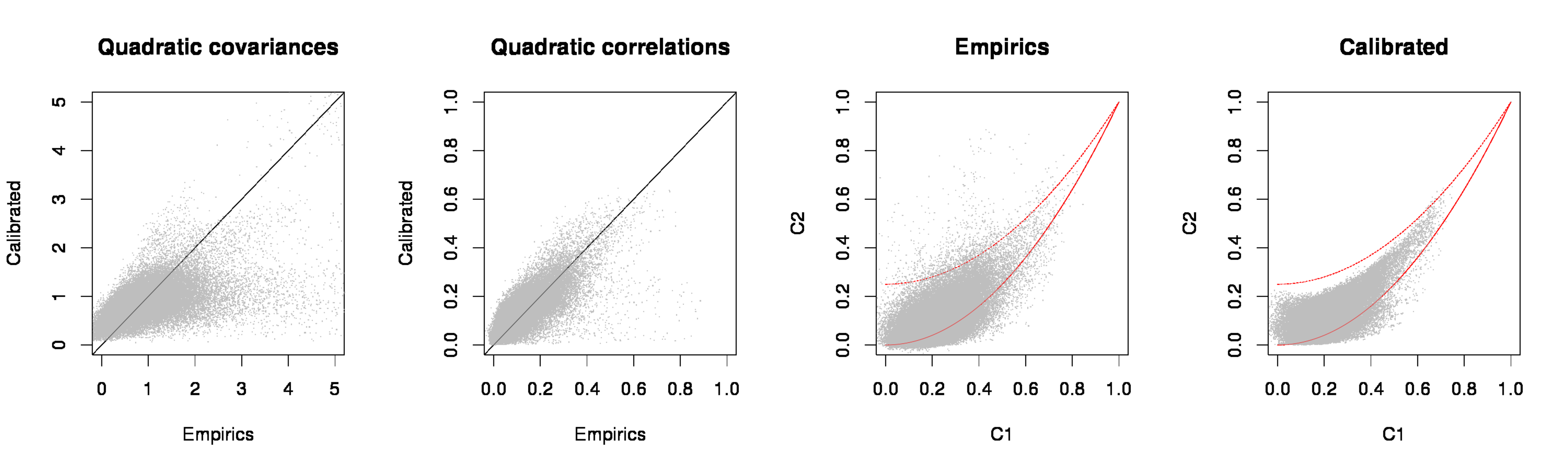}}\\
    \subfigure[2005--2009]{\includegraphics[scale=.2,trim=750 50 0 50,clip]{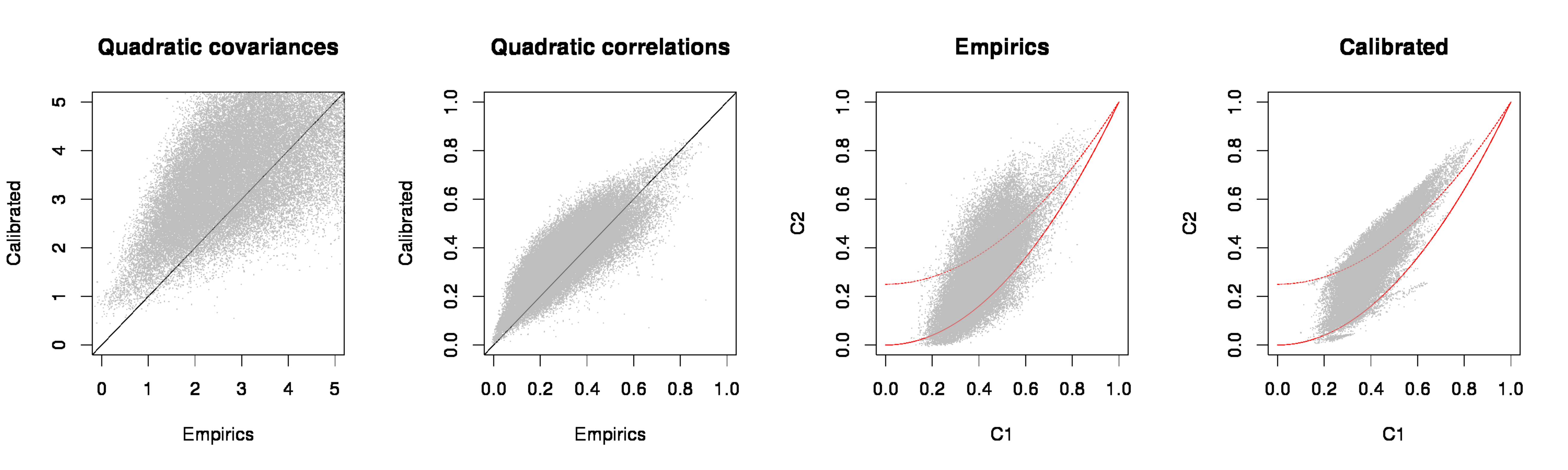}}\\
    \subfigure[2009--2012]{\includegraphics[scale=.2,trim=750 50 0 50,clip]{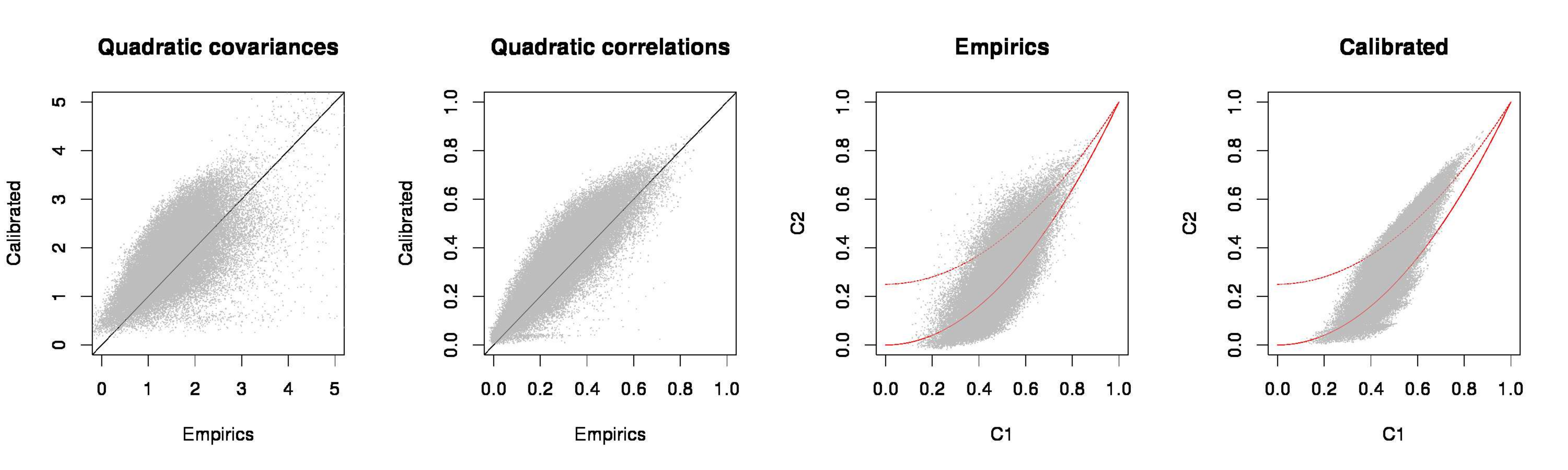}}
    \caption{\textbf{Left: }   calibrated vs sample quadratic correlations. 
             \textbf{Middle: } sample quadratic correlations vs sample linear correlations;
             \textbf{Right: }  calibrated quadratic correlations vs calibrated linear correlations.
             Two benchmark curves are added in red: the Gaussian case (lower curve) and the Student case with $\nu=5$ d.o.f.\ (upper curve).
             \label{fig:cal_quad}}
\end{figure}

The quadratic correlations can be explicitly computed from the model definition, and write:
\begin{align}\nonumber
\esp{\ret_i^2\ret_j^2}  &=\sum_{k\ell}\left(\Wei_{ki}^2\Wei_{\ell j}^2+2\Wei_{ki}\Wei_{kj}\Wei_{\ell i}\Wei_{\ell j}\right)\Phi_0(A_{k0},A_{\ell0};2)
\Big(\tfrac{1}{3}\cdot 3\cdot \Phi_{\text{G}}(s_{k},s_{\ell};2)\Big)^{\delta_{k\ell}} \\\nonumber
                        &+(1+2\delta_{ij})\Big(1-\sum_\ell\Wei_{\ell i}^2\Big)\sum_{k}\Wei_{kj}^2\Phi_0(A_{k0},B_{i0};2)\\\nonumber
                        &+(1+2\delta_{ij})\Big(1-\sum_\ell\Wei_{\ell j}^2\Big)\sum_{k}\Wei_{ki}^2\Phi_0(A_{k0},B_{j0};2)\\\label{eq:quad_cor_model}
                        &+\Big(1-\sum_\ell\Wei_{\ell i}^2\Big)\Big(1-\sum_\ell\Wei_{\ell j}^2\Big)\Phi_0(B_{i0},B_{j0};2)\Big(3\Phi_{\text{G}}(\widetilde s_{i} ,\widetilde s_{j} ;2)\Big)^{\delta_{ij}},
\end{align}
where $\Phi_0$ is defined in Appendix~\ref{app:calibVol}.
When all parameters $A,B,s$ are zero, the prediction for Gaussian factors and residuals is retrieved: $\esp{\ret_i^2\ret_j^2}\equiv 1+2\esp{\ret_i\ret_j}^2$.
We illustrate in the left panel of Fig.~\ref{fig:cal_quad} a scatter plot of 
the left-hand side (calibrated) versus the right-hand side (empirical) of Eq.~\eqref{eq:quad_cor_model}, 
for all periods. They show a good agreement of model and sample quadratic correlations.
Furthermore, the middle and right panels of the same figure illustrate the fact that the
pairs of stock returns cannot be described by a bivariate Student distribution, 
for which a regular curve should be observed instead of the scattered cloud 
in the plane of quadratic vs linear correlations. This conclusion was already reached in \cite{chicheportiche2012joint}, and is 
made precise by the present nested factor model.%
\footnote{
Notice that the choice of $p$ in the estimation procedure of the parameters $B_{i0}$ and $\widetilde s_{i}$ is important here.
Estimation biases and errors are in practice different for low moments $p\approx 0.2$ or high moments $p\approx 2$.
Obviously, best fits for the quadratic correlations are obtained with $p=2$ since in this case 
the same quantities appear in Eq.~\eqref{eq:quad_cor_model} and in the loss function \eqref{eq:loss_B}.
}

\subsubsection{Copulas: medial point and diagonals}

The middle point $\cop(\frac12,\frac12)$ of the copula, which was shown on Fig.~\ref{fig:emp_cop} to be incompatible with any elliptical prediction,
is also very well captured by our model, with no further ingredients. 
Although an analytical expression relating  $\cop(\frac12,\frac12)$ to the model parameters is out of reach, 
it is possible to reproduce its predicted value by simulating long time series according to the model with estimated parameters.%
\footnote{The non-Gaussian series of log-volatility $\Omega_{t0}$ is generated as independent realizations of a Beta distribution
whose coefficients are determined so that the first four moments match those of $\Omega_0$.
This class of distributions allows for negative kurtosis.
It is known that the realizations of volatility exhibit strong persistence, 
a characteristic that our simulated series do not reproduce.
This however does not generate a bias in the obtained coefficients, but rather makes them ``not noisy enough''.
}
The results are in remarkable agreement with the data (see Fig.~\ref{fig:emp_cop}), and emphasize the capacity of our non-Gaussian factor model to cope with the
non-trivial behavior of the medial point of the copula.

\begin{sidewaysfigure}%[p]
\center
\subfigure[Empirical]{ \includegraphics[scale=.5]{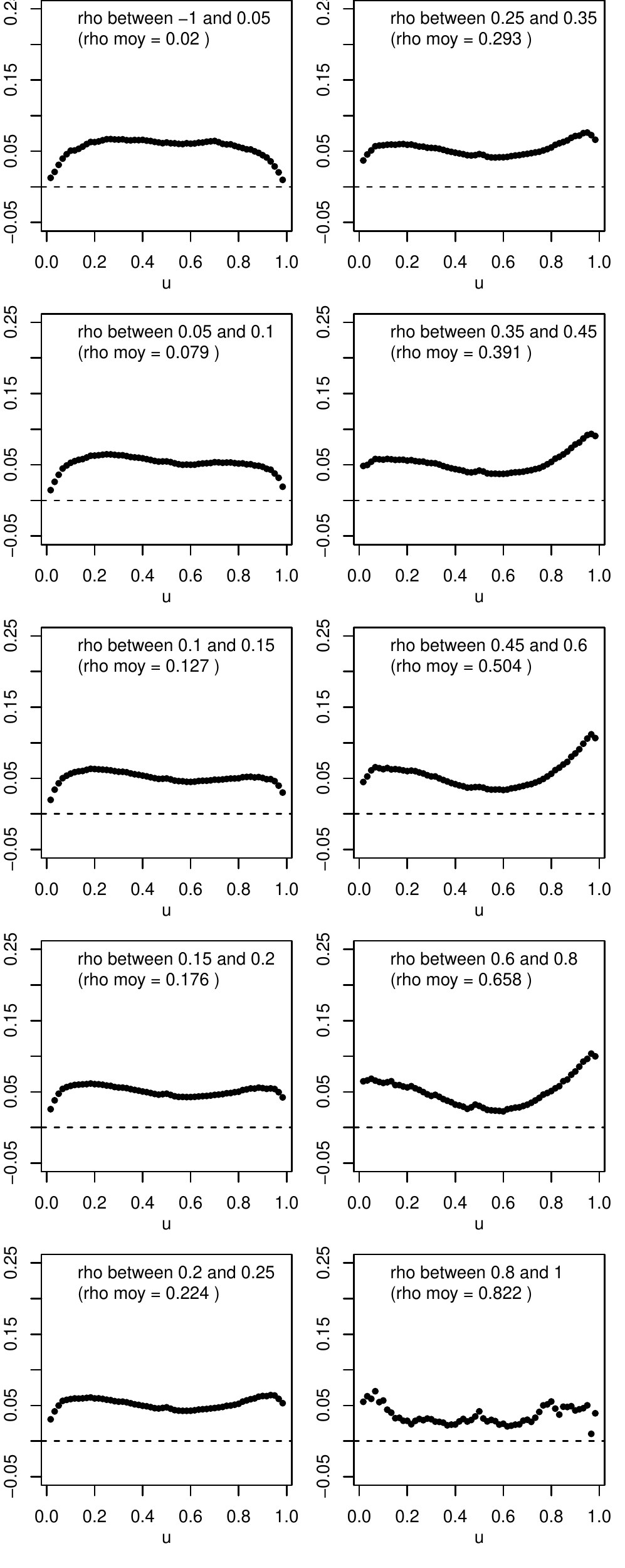}}
\subfigure[Calibrated]{\includegraphics[scale=.5]{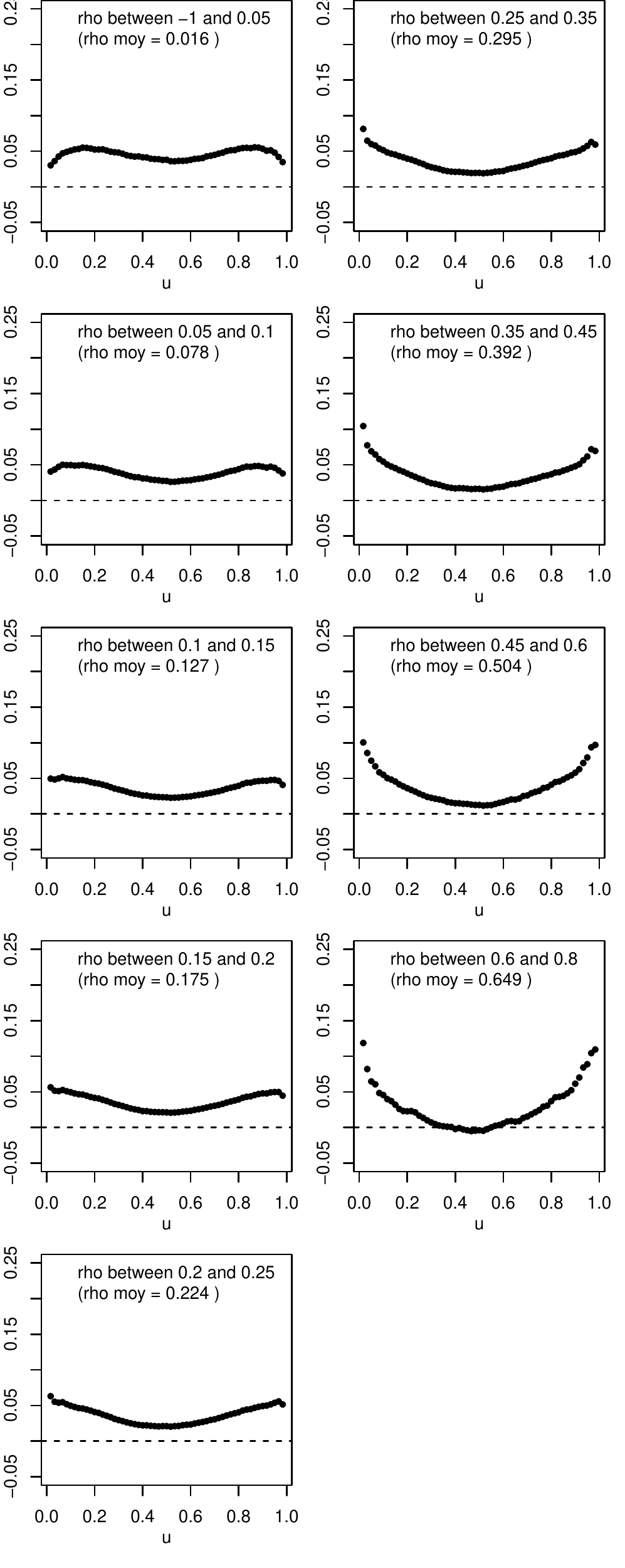}}
\hfill
\subfigure[Empirical]{ \includegraphics[scale=.5]{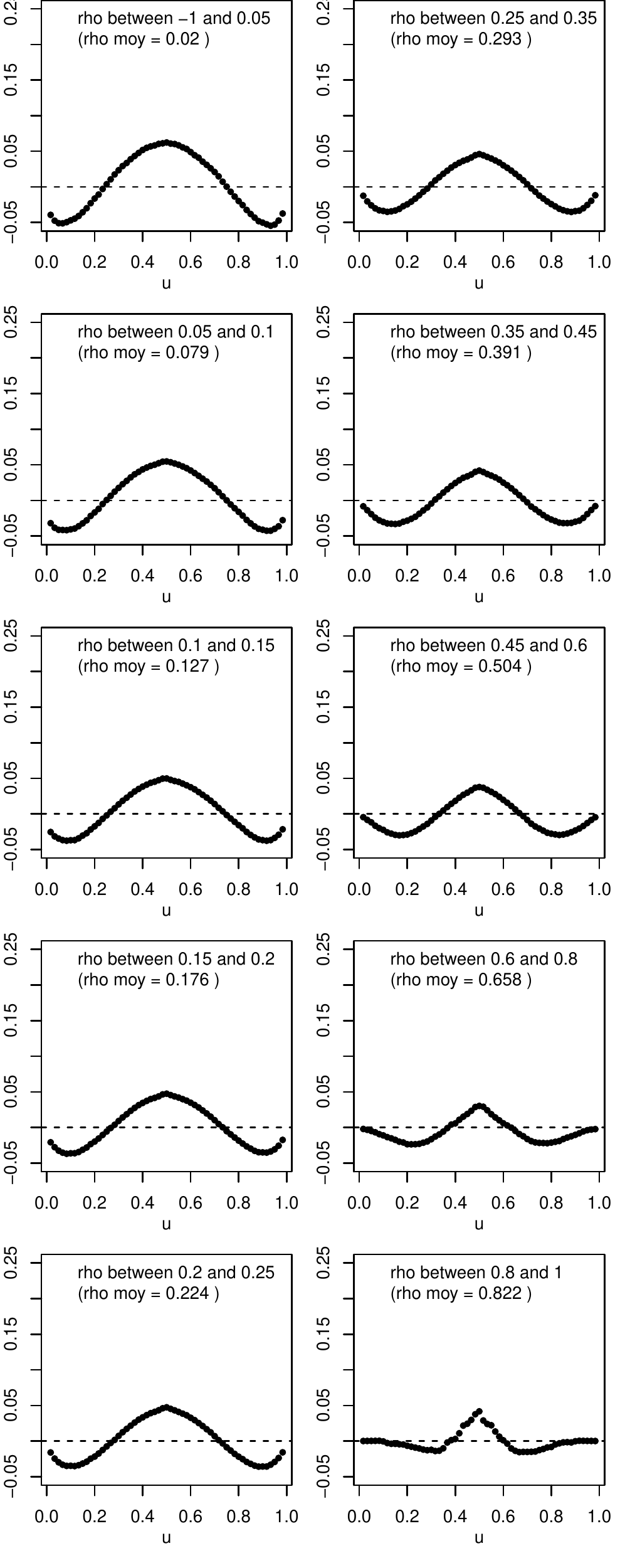}}
\subfigure[Calibrated]{\includegraphics[scale=.5]{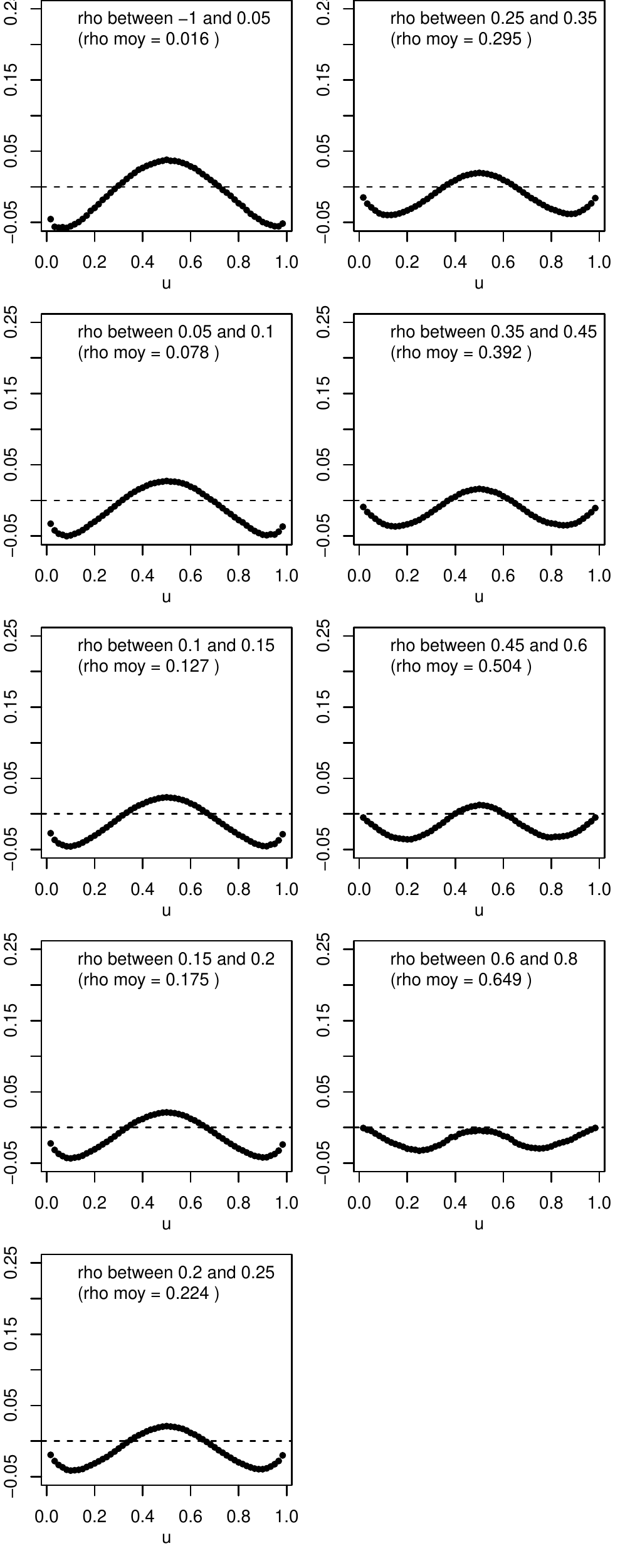}}
\caption{2000--2004, diagonal (left) and anti-diagonal (right) of the copula, as in Eq.~\eqref{eq:diagcopdef}.}
\label{fig:side_cop20002004}
\end{sidewaysfigure}

This is confirmed and in fact strengthened by the analysis of the bivariate copulas along the whole diagonals.
Fig.~\ref{fig:side_cop20002004} 
compares empirically measured and model-predicted values of the quantities
\begin{equation}\label{eq:diagcopdef}
    \Delta_{\scriptscriptstyle \text{d}}(p)=\frac{\cop(p,p)-\cop[\text{G}](p,p)}{p\,(1-p)}
    \quad\text{and}\quad
    \Delta_{\scriptscriptstyle \text{a}}(p)=\frac{\cop(p,1-p)-\cop[\text{G}](p,1-p)}{p\,(1-p)}
\end{equation}
versus $p$, for several values of the linear correlation over 2000--2004 (similar plots for other periods are produced in \cite{chicheportiche2013phd}).
A direct visual comparison reveals that the main non-trivial qualitative features of the empirical diagonal copulas
are well reproduced by our model. For example, the evolution of the concavity as $\rho$ changes, the behavior in the tails, 
and the medial-point behavior as discussed above. Plots of similar quality have been obtained for other sub-periods as well.

One may note however that the asymmetry $u\leftrightarrow 1-u$, visible in the graphs, is {\it not} reproduced by our fully symmetric model, and would require accounting for the leverage
effect, i.e.\ cross-correlations between the linear factors and residuals $f_k, e_j$, and the volatility factors and residuals $\Omega_0, \omega_k, 
\widetilde \omega_j$.

Before proceeding to the out-of-sample evaluation of the model in Sect.~\ref{sec:stability},
we briefly present how the model can be improved (though marginally)
by introducing a second volatility driver $\Omega_{1}$, as suggested by the spectral analysis of Sect.~\ref{sec:MFspectral}.

\subsection{A second volatility driver}
The spectral analysis of the factor and residual absolute correlations has revealed
that there exists a small but significant second mode of volatility.
The model in Eq.~\eqref{eq:model0} can be improved accordingly in order to account for this additional source of collective amplitude fluctuations, see 
Eq.~\eqref{eq:model_fkej_0} above.

The whole estimation procedure runs identically.
However, for the determination of the parameters $A_{k0}$ and $A_{k1}$, 
the reduced number of observations ($M(M-1)/2$ factor-factor correlations, times 8 values of $p$)
provides only a low resolution, and the minimization program does not succeed in distinguishing the two volatility drivers:
it outputs an hybrid where both $\Omega_0$ and $\Omega_{1}$ contribute to the same mode.
In order to break the degeneracy and ``orthogonalize'' the modes, we add an overlap term $\left(\sum_k A_{k0}A_{k1}\right)^2$ 
in the cost function Eq.~\eqref{eq:costfct_A}.

\begin{figure}
    \center
    \subfigure[$A_{k0}$ and $A_{k1}$]{\label{fig:0004_2_A}\includegraphics[scale=.5,trim=  0 140 0 0,clip]{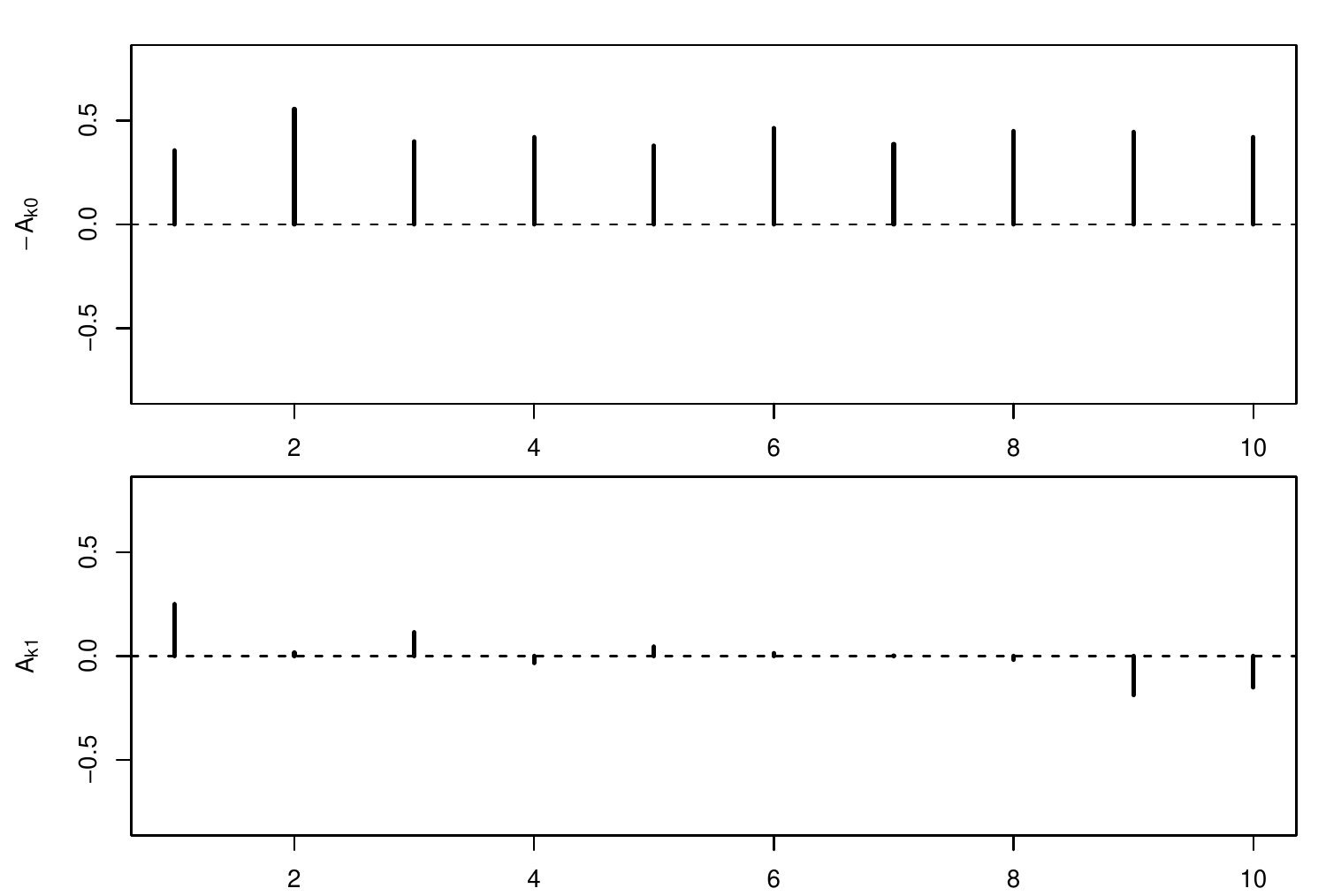}}%\hspace{.5cm}
    \subfigure[$B_{j0}$ and $B_{j1}$]{\label{fig:0004_2_B}\includegraphics[scale=.5,trim=  0 140 0 0,clip]{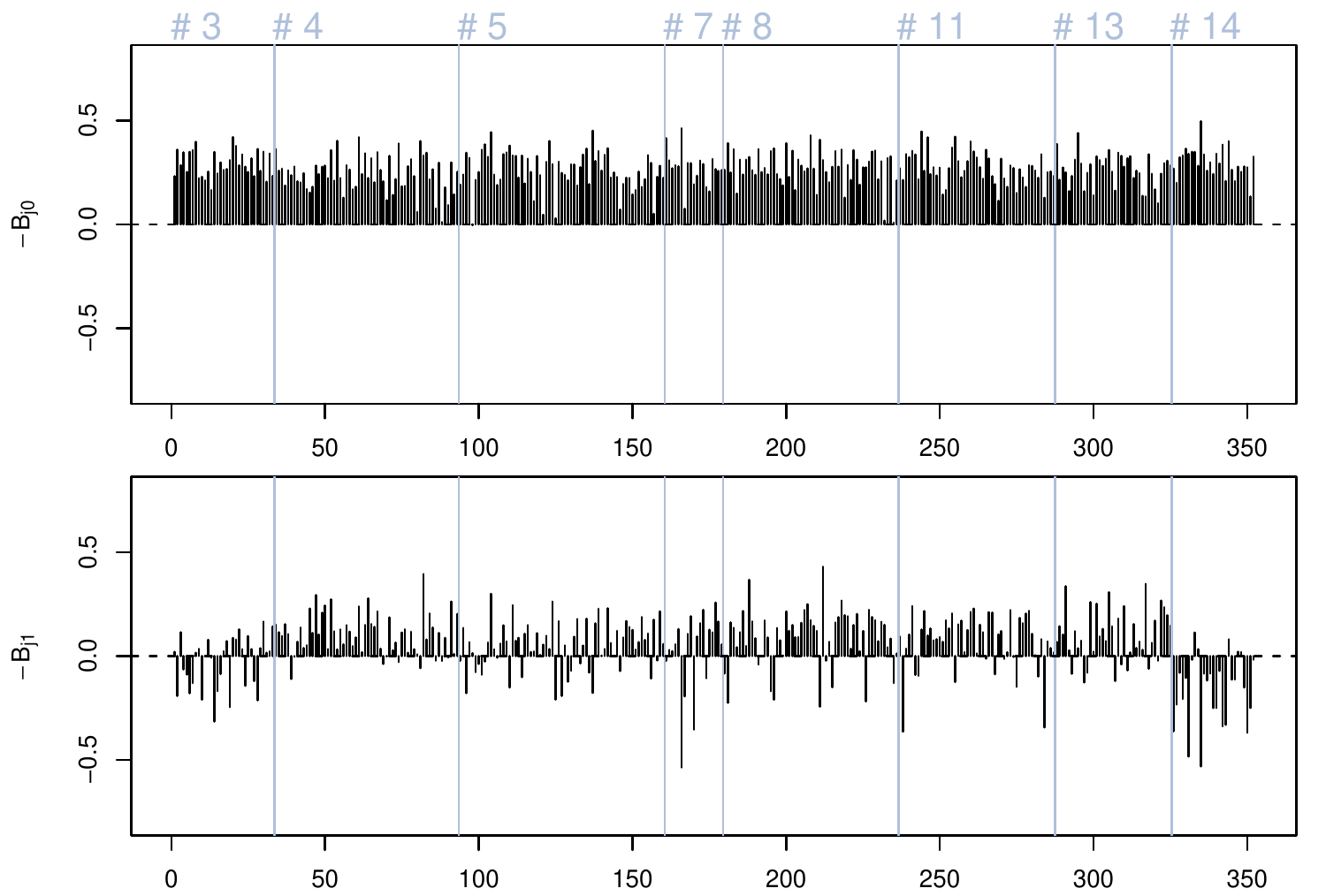}} 
    \vfill
    \subfigure[Index amplitudes reproduced with two volatility drivers $\Omega_0$ and $\Omega_1$, to be compared with Fig.~\ref{fig:firstfact1mode}]{\label{fig:0004_2_O}\includegraphics[scale=1,trim=220 0 0 0,clip]{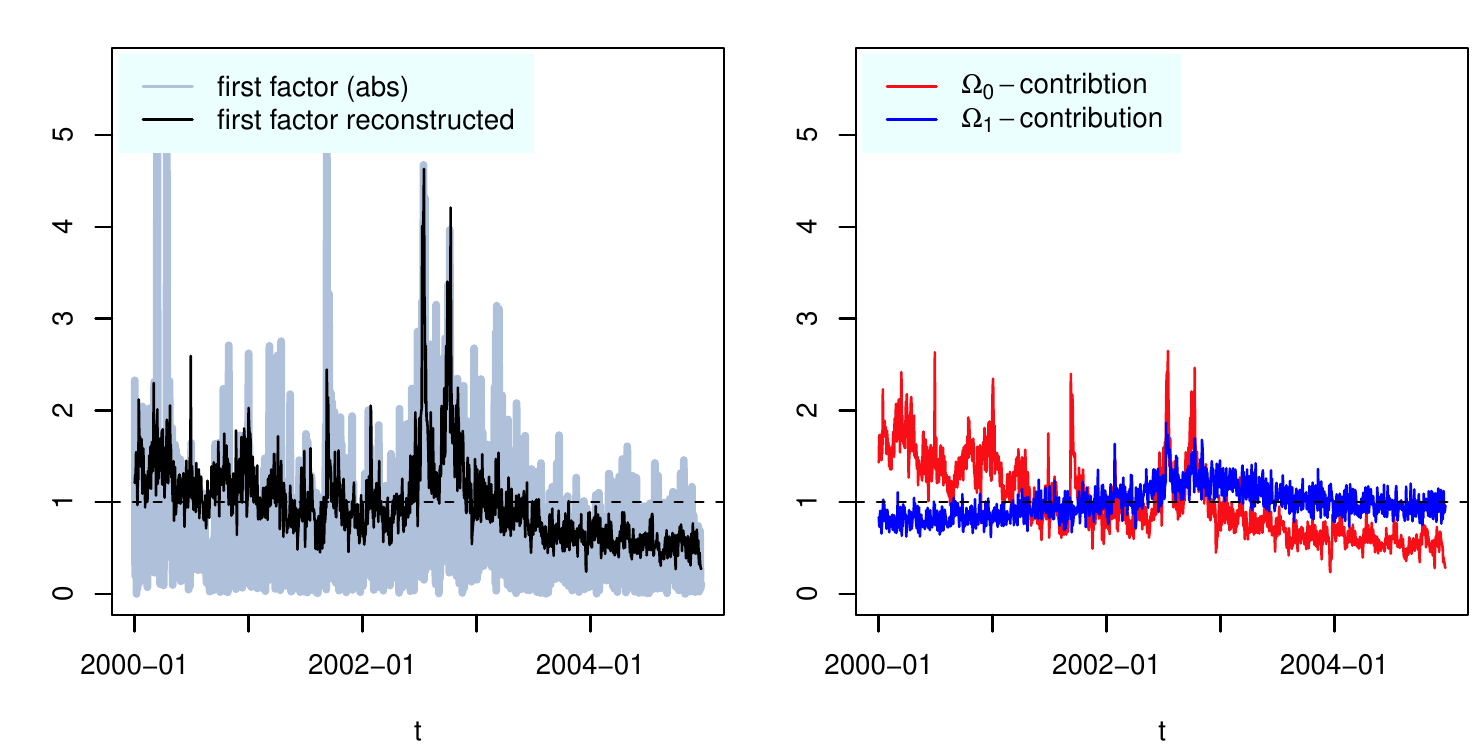}}
    \caption{Two volatility modes  $\Omega_0$ and $\Omega_1$. $M=10$, 2000--2004}\label{fig:2modes0004}
\end{figure}

As an example, we report in Fig.~\ref{fig:2modes0004} the results for the period 2000--2004.
As expected, the parameters $A_{k0}$ and $A_{k1}$ are very close to the first two 
eigenvectors of the factor-factor ``log-abs'' correlation matrix, and the parameters $B_{j0}$, $B_{j1}$ look like the first two eigenvectors 
of the residual-residual  matrix. Clearly, taking this additional second volatility driver into consideration
improves the theoretical description of the returns.
We illustrate this on Fig.~\ref{fig:0004_2_O} where we show how $\Omega_0$ and $\Omega_{1}$ contribute  to the volatility of the market mode of linear correlation, $f_1$.

\section{Out-of-sample analysis}\label{sec:stability}

All the results presented above are ``in-sample'', 
in the sense that we have shown the predicted dependence coefficients with estimated parameters on a period
and compared them to the realized coefficients in that same period.
The ultimate test for a model that aims at describing joint financial returns (and more generally of any risk model), 
is to improve ``out-of-sample'' predictions, i.e.\ use a model calibrated on a period to predict some quantity in a \emph{subsequent} period.

We will test the different models through the predicted correlation matrix. For the (linear) correlation matrix of the returns themselves, this 
has been the subject of many papers in the literature already, 
see \cite{laloux1999noise,laloux2000random,ledoit2004well,potters2005financial,tumminello2007shrinkage,elkaroui2010high,bartz2012directional,potters2009financial}. 
Even if this is not the primary aim of the present study, we will first test the ability of our linear factor model to 
correctly predict the out-of-sample risk of optimal linear portfolios. We will then turn to the case of a portfolio of non-linear assets (absolute values), which 
has not been considered so far in this context (to the best of our knowledge). We will show that 
our ``dominant volatility mode'' framework outperforms other natural models for predicting out-of-sample risk. 

We will consider a long period 2000--2009 on which we perform an In-sample/Out-of-sample analysis over sliding windows ($N=262$ returns series are kept, see Tab.~\ref{tab:sectors}).
We rely on the procedure introduced by \cite{potters2009financial}, that we reproduce for convenience in Appendix~\ref{app:ISOS}.

\subsection{Linear correlations}

We first revisit the standard Markowitz problem, attempting to minimize the out-of-sample risk of an optimal portfolio constructed using as input different 
correlation matrices:
 \begin{itemize}
 \item{Empirical: }     the in-sample raw correlation matrix, $$\displaystyle \rod[1]_\text{Emp}(\tau)=\frac{1}{T^{\scriptscriptstyle \text{IS}}}\sum_{t'=\tau-T^{\scriptscriptstyle \text{IS}}}^{\tau-1}\Ret_{t'\cdot}\cdot\Ret_{t'\cdot};$$
 \item{Ledoit-Wolf \citep{ledoit2004well}:}  the convex combination 
                        \begin{align*}
                        \rod[1]_\text{LW}(\tau)&=\alpha \rod[1]_\text{Emp}(\tau)+(1-\alpha)\overline{\rod[1]}(\tau), \\
       \text{where}\quad\overline{\rod[1]}_{ij}&=\begin{cases}\displaystyle 1&,i=j\\\displaystyle \frac{1}{N(N-1)/2}\sum_{i',j'<i'}[\rod[1]_\text{Emp}]_{i'j'}&,i\neq j\end{cases}.
                        \end{align*}
                        This corresponds to a ``shrinkage'' of the noisy sample correlation matrix toward its rank-one approximation.
 \item{Clipped: }       $\rod[1]_\text{Clip}(\tau)$ retaining only the $M$ eigenmodes of $\rod[1]_\text{Emp}(\tau)$ with largest eigenvalues, 
                        and adjusting all remaining eigenvalues to $(N-\sum_{i=1}^M\lambda_i)/(N-M)$ in order to conserve the trace;
%\item{Clipped: }       the PCA solution of Eq.~\eqref{eq:X_PCA}, $\rod[1]_\text{Clip}(\tau)=\Wei_\text{PCA}(\tau)^\dagger\Wei_\text{PCA}(\tau)$ (for off-diagonal elements), 
%                       for several values of the number of retained modes $M$ with largest eigenvalues;
 \item{MultiFactor: }   the improved solution of Eq.~\eqref{eq:offdiag_content_apx}, $\rod[1]_\text{F}(\tau)=\Wei_\text{F}(\tau)^\dagger\Wei_\text{F}(\tau)$ (for off-diagonal elements), 
                        for several values of the number of factors $M$.
 \end{itemize}
 All these scenarii can furthermore be compared to the benchmark of a full-rank pure noise Wishart correlation matrix.
 In this case, Random Matrix Theory predicts the values of the average in-sample and out-of-sample risks,
 in the limit of large matrices with quality factor $q=N/T^{\scriptscriptstyle \text{IS}}$ \citep{potters2009financial}:
 \[
    \vev{\mathcal{R}^2_\text{RMT}}_\text{IS}=\mathcal{R}^2_\text{True} \cdot (1-q)
    \qquad\text{and}\qquad
    \vev{\mathcal{R}^2_\text{RMT}}_\text{OS}=\mathcal{R}^2_\text{True}  \,/\,(1-q).
 \]
 Moreover, the true risk (i.e.\ the value of $\mathcal{R}^2$ when the optimal weights are determined using the correlation matrix 
 of the process that generates the realized returns $\Ret_{\cdot i}/\sigma_i^{\scriptscriptstyle \text{IS}}$) 
 can be shown to be $\mathcal{R}^2_\text{True}=1$ with the definition \eqref{eq:def_risk}.
 
\begin{figure}
    \center
    \includegraphics[scale=0.53]{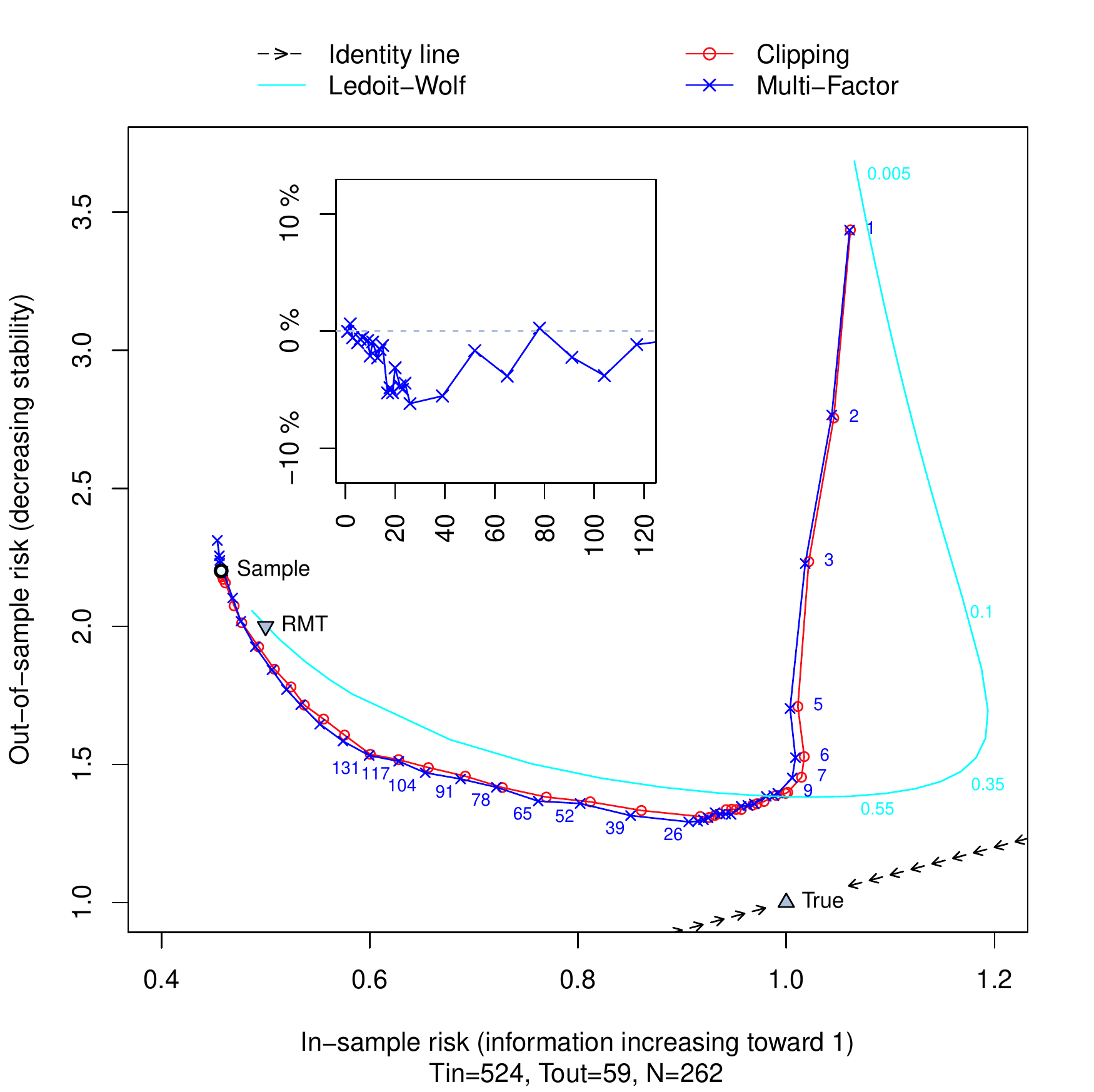}
    \caption{Linear correlations: Out-of-sample risk vs In-sample risk, defined in Eq.~\eqref{eq:def_risk}
    and averaged over sliding windows in 2000--2009, 
    for three cleaning schemes: eigenvalue clipping (red circles) and calibrated multi-factor model (blue crosses), 
    both with $M$ linear factors, as well as Ledoit-Wolf shrinkage (cyan line).
    Inset : the Out-of-sample risk is lowered by more than $5\%$ with respect to ``Clipping'', when $M\approx 24$.}
    % \caption{Linear correlations: In-sample risk (lower curves) and out-of-sample risk (upper curves) defined in Eq.~\eqref{eq:def_risk}
    % and averaged over sliding windows in 2000--2009, 
    % for three cleaning schemes: eigenvalue clipping (red circles) and calibrated multi-factor model (blue crosses), 
    % both with $M=\alpha N$ linear factors, as well as Ledoit-Wolf shrinkage (cyan stars).
    % Inset : the out-of-sample risk is lowered by more than $5\%$ with respect to ``Clipping''; 
    % the location of the minimum coincides with that of minimum risk ($\alpha\approx 0.1$).}
    \label{fig:ISOS_lin}
\end{figure}

We show graphically the results of the testing procedure on Fig.~\ref{fig:ISOS_lin}:
In-sample and Out-of-sample average risks of every cleaning scheme are plotted parametrically with a control parameter $\alpha$ (equal to $M/N$ for Clipping and MultiFact),
 where averages are performed over the sliding windows $(\tau)$.

When only a very reduced number of factors ($M\approx 1,2,3$) is kept, eigenvalue clipping performs better (although quite bad), 
and similarly when keeping also the very last modes:  this is because the linear factors are only good when the eigenmodes are statistically significant, on the left and right of the RMT noise bulk.
In the limit $\alpha\to 1$ (i.e.\ $M=N$) all cleaning schemes collapse to the risk values associated with
the raw ``Sample'' correlation matrix. The benchmark RMT prediction is shown for reference.

In the intermediate regime, the Out-of-sample risk is minimal because the marginal gain in signal is higher than the marginal risk increase due to added ``false information''.
In this case, it turns out that the ``factor model'' procedure worked out in Appendix~\ref{app:PCA} provides an improved determination of average Out-of-sample risk.
In fact, the inset of Fig.~\ref{fig:ISOS_lin} shows that the relative gain
\begin{equation}\label{eq:rel_gain}
    [\vev{\mathcal{R}^2_{\text{Clip}}}-\vev{\mathcal{R}^2_{\text{MF}}}]/[\vev{\mathcal{R}^2_{\text{Clip}}}-\mathcal{R}^2_{\text{True}}]
\end{equation}
 can reach up to 6--7\%, while not dramatically increasing over-fitting: the In-sample risk is only slightly artificially lowered.
 
 Therefore, although our aim was to set up a model that would describe faithfully the non-linear dependences between stocks, we find that the first step
 of our procedure, namely the calibration of a factor model to capture the {\it linear} correlations, leads to the best cleaning procedure so far (at least
 for the out of sample risk criterion we use here).
 
\subsection{Absolute correlations}
\begin{figure}
    \center
    \includegraphics[scale=0.53]{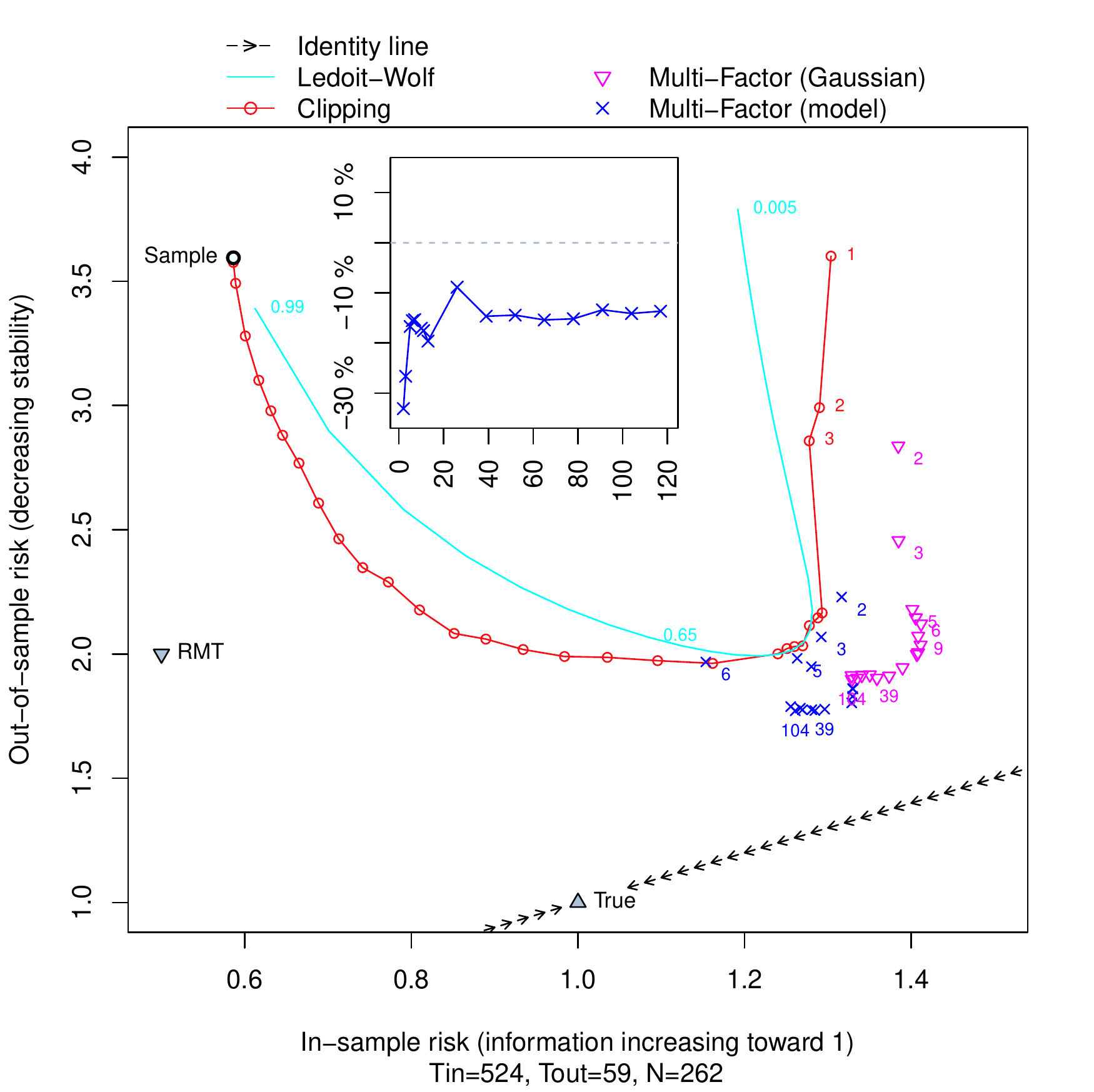}
    \caption{Absolute correlations: Out-of-sample risk vs In-sample risk, defined in Eq.~\eqref{eq:def_risk}
    and averaged over sliding windows in 2000--2009, 
    for three cleaning schemes: eigenvalue clipping of $M$ modes of quadratic correlation (red circles),
    Gaussian multi-factor (magenta triangles) and calibrated multi-factor model (blue crosses), 
    both with $M$ linear factors.
             The role of volatility dependences is elicited by the better performance of the non-Gaussian
             multi-factor level over the Gaussian multi-factor cleaning scheme: the relative risk difference \eqref{eq:overperfRatio} is shown in the inset, and is always negative.}
    % \caption{Absolute correlations: In-sample risk (lower curves) and out-of-sample risk (upper curves) defined in Eq.~\eqref{eq:def_risk}
    % and averaged over sliding windows in 2000--2009, 
    % for three cleaning schemes: eigenvalue clipping of $M=\alpha N$ modes of quadratic correlation (red circles),
    % Gaussian multi-factor (blue triangles) and calibrated multi-factor model (blue crosses), 
    % both with $M=\alpha N$ linear factors.
             % The role of volatility dependences is put forward by the better performance of the non-Gaussian
             % multi-factor level over the Gaussian multi-factor cleaning scheme: the relative risk difference \eqref{eq:overperfRatio} is shown in the inset.}
    \label{fig:ISOS_abs}
\end{figure}

We now turn to the core property of our model: its ability to capture non-linear dependences.
We have already shown that the model is able to reproduce, after calibration, several empirically observed quantities
like the copula,
and want now to perform an out-of-sample assessment of the volatility-driven dependence in the absolute correlations.
The definitions of the gain predictor $\vect{g}$ and the risk measure $\mathcal{R}^2(\tau)$ are identical
to Eqs.~\eqref{eq:markowitz_g} and \eqref{eq:def_risk} respectively (see Appendix~\ref{app:ISOS}), with now 
\[
    \mat{Y}_{ti} = \frac{|\Ret_{ti}|-\vev{|\Ret_{ti}|}_t}{\sqrt{\Vev{\left(|\Ret_{ti}|-\vev{|\Ret_{ti}|}_t\right)^2}_t}}.
\]
The different cleaning schemes considered are:
 \begin{itemize}
 \item{Empirical: }          the in-sample raw correlation matrix, 
                             $$\displaystyle \rod[\text{a}]_\text{Emp}(\tau)=\frac{1}{T^{\scriptscriptstyle \text{IS}}}
                               \sum_{t'=\tau-T^{\scriptscriptstyle \text{IS}}}^{\tau-1}\mat{Y}_{t'\cdot}\cdot\mat{Y}_{t'\cdot};$$
 \item{Ledoit-Wolf \citep{ledoit2004well}:}  the convex combination 
                        \begin{align*}
                        \rod[a]_\text{LW}(\tau)&=\alpha \rod[a]_\text{Emp}(\tau)+(1-\alpha)\overline{\rod[a]}(\tau),
                        \end{align*}
                        similarly to the linear case. This corresponds to a ``shrinkage'' of the noisy in-sample cor-abs matrix toward its rank-one approximation.
 \item{Clipped: }       $\rod[a]_\text{Clip}(\tau)$ retaining only the $M$ eigenmodes of $\rod[a]_\text{Emp}(\tau)$ with largest eigenvalues, 
                        and adjusting all remaining eigenvalues to $(N-\sum_{i=1}^M\lambda_i)/(N-M)$ in order to conserve the trace;
%\item{Clipped: }            $\rod[\text{a}]_\text{Clip}(\tau) 
%                            keeping the $M$ eigenmodes of $\rod[\text{a}]_\text{Emp}$ with largest eigenvalues; %***que fais tu de la diagonale ???***
 \item{Gaussian factors: }   the Gaussian prediction $\rod[\text{a}]_\text{FG}(\tau)$ obtained as the sample absolute correlations of long time series simulated
                             according to the $M$-factor model where all volatility parameters $A$, $B$ and $s$ are set to 0.
 \item{Multifactor (model):} the model prediction $\rod[\text{a}]_\text{FnG}(\tau)$ obtained as the sample absolute correlations of long time series simulated (an analytic expression of absolute correlations is out of reach)
                             according to the $M$-factor model with one volatility mode.
 \end{itemize}
Notice that the meaning of $M$ is not comparable in all cleaning schemes:
while for the ``clipped eigenvalues'' it corresponds to the number of relevant modes in the 
matrix of \emph{absolute correlations}, 
for the multi-factor models it instead counts the the number of \emph{linear} factors.
This can be seen immediately on Fig.~\ref{fig:ISOS_abs}, where the red curve corresponding to ``Clipping''
has the usual U-shape, while the blue and magenta points corresponding to ``multi-factor'' saturate as $M$ increases above $\approx 30$,
a threshold above which letting additional linear factors barely affects the volatility dependences.

More importantly, this figure shows that multi-factor models offer a better optimal Out-of-sample risk together with less In-Sample over-fitting.
The role of volatility dependences is  revealed by the much better performance of the final non-Gaussian
multi-factor level over the Gaussian multi-factor cleaning scheme.
This is emphasized in the inset plot of Fig.~\ref{fig:ISOS_abs} representing the over-performance ratio
\begin{equation}\label{eq:overperfRatio}
        [\vev{\mathcal{R}^2_{\text{FnG}}}-\vev{\mathcal{R}^2_{\text{FG}}}]/[\vev{\mathcal{R}^2_{\text{FnG}}}-\mathcal{R}^2_{\text{True}}].
\end{equation}
%with a reference $\mathcal{R}^2_\text{ref}=1.5$ 
%(arbitrarily chosen between the highest average In-sample risk and the lowest Out-of-sample risk),
%since the true risk for absolute returns is not 1.
The non-Gaussian model performs always better than the Gaussian model.

\subsection{How many factors should be kept?}

The number $M$ of linear factors in the description \eqref{eq:MODEL} is an important input of the model.
The intuition that statistical factors are somewhat related to economic sectors does not stand
the identification of algebraic modes of fluctuations to sectorial or other macro-economic factors,
beyond the first two or three modes. Still, even if there is no one-to-one identification, 
the number of sectors can be regarded as a reasonable prior for $M$.
In our calibration, we have retained $M=10$ factors corresponding to the number of Bloomberg sectors plus one,
with satisfactory results at reproducing the main empirical stylized facts.

A more convincing determination of $M$ is reached by reconsidering the above results on the In-Sample/Out-of-Sample risk 
test for linear portfolios. From a general standpoint, we know that there must exists an optimal number of parameters,
for which the model fits reasonably the data and avoids over-fitting, i.e.\ is stable when applied Out-of-Sample.
Adjusting the ratio $\alpha=M/N$ allowed us to find an optimal configuration
where the Out-of-Sample risk is minimized while the In-Sample risk is not artificially lowered.
A value of $\alpha \approx 0.1$ is found to be optimal for the standard (linear) Markowitz problem,
while the the risk associated with absolute returns is lowered by our non-Gaussian factor model whatever the number of linear modes kept. 
This suggests that the optimal number of factor to be kept is $M\approx 24$ for the $262$ stocks considered here.
    
\section{Conclusions}

Finding a faithful mathematical representation of the multivariate distribution of stock returns in a given market 
is one of the unsolved problems in quantitative finance. Copula-based research efforts have investigated a host of 
different possibilities, with disappointing results --- both because the proposed copulas are not able to grasp the 
clear ``stylized facts'' evidenced by empirical studies, and because most of these copulas lack financial motivation and
intuition. Among these stylized facts, two are particularly striking \citep{cizeau2001correlation,allez2011individual,chicheportiche2012joint}:
\begin{itemize}
\item The market factor (index) volatility is strongly correlated with the volatility of residuals, even when the index and residuals
are by construction uncorrelated. This naively suggests a multiplicative structure for stock returns, schematically as $r = \sigma (\beta f + e)$.
\item However, empirical copulas are incompatible with the above multiplicative (pseudo-elliptical) structure. In particular, medial-points 
of bivariate copulas have a non-trivial dependence on the linear correlation, that rules out this family of dependence, except for very 
highly correlated pairs of stocks.
\end{itemize}

The aim of our work was to propose a natural framework to account for all the empirical properties of the multivariate distribution of stock returns.
We defined a ``nested factor model'', where the linear factors part is standard (apart from the calibration procedure), but where the log-volatility of the
linear factors {\it and} of the residuals are themselves endowed with a factor structure and residuals. We proposed a calibration procedure to estimate 
these log-vol factors and the residuals. We found that whereas the number of relevant linear factors is relatively large ($10$ or more), only two or three log-vol
factors emerge in our analysis of the data. In fact, a minimal model where only {\it one} log-vol factor is considered is already very satisfactory, as it accurately reproduces the 
properties of bivariate copulas, in particular the subtle medial-point properties mentioned above. We have tested the ability of our model to predict Out-of-Sample the
risk of non-linear portfolios, and found that it performs significantly better than all other schemes that we could think of. 

The nested factor structure of the model makes it difficult to write down explicitly the corresponding copulas. This illustrates why a formal approach to 
multivariate copulas is doomed to fail: copulas are not necessarily the natural language in which the specificities of financial markets can be elicited.  

There are many avenues of research suggested by the present study. First, it would be interesting to check that other stock markets (EU, UK, JP) lead 
to the same conclusions, as we believe they will. Second, a joint analysis of the multivariate properties of stock returns and implied volatilities in 
the corresponding option markets would be highly worthwhile. Third, as we have pointed out above, our model is at this stage purely static, in the sense that
we have not specified the dynamics of volatility modes and residuals. 
\cite{morales2013dependency} propose a first step toward the integration of cross-sectional dependences and dynamic (scaling) properties of financial time series: 
they show that multifractality and cross-correlations are much related and suggest a hierarchical construction of stock dependences 
able to account at the same time for the multi-scaling.
This is a very rich subject, since all these objects are expected to show long-range
temporal dependence, leverage effects and possibly lagged cross correlations between them. Note that, as we already emphasized, the dominant volatility factor is
{\it not} the volatility of the market factor. The decomposition of the well known index leverage effect into its various components, and the consequences for 
index options pricing and VIX, is a very natural issue to investigate first.

\subsection*{Acknowledgements} This work has benefitted from many insightful comments from
J.~Bun, S.~Ciliberti, B.~Durin, P.~Horvai, M.~Potters, L.~Laloux, E.~S\'eri\'e and G.~Simon.

\appendix

\section{Factor models and PCA}\label{app:PCA}

The Principal Components Analysis (PCA) relies on the spectral decomposition
of the covariance matrix. It is related to, but different from the logic of the linear factor model. In fact, as we show below, 
the PCA provides a starting point for the identification of the weights $\Wei_{\text{F}}$ of the factors.
    
    The diagonalization of the sample correlation matrix yields
    \[
        \frac{1}{T}\Ret^\dagger \Ret = \mat{V}\Lambda \mat{V}^\dagger
    \]
    where $\Lambda$ is the diagonal matrix of eigenvalues, and the columns of $\mat{V}$ are the corresponding eigenvectors.
    Hence, there always exist (linearly orthogonal) factor series $\widetilde{\mat{F}}$ such that the return series $\Ret$ can be decomposed as
    \begin{equation}\label{eq:PCA_sol}
        \Ret=\widetilde{\mat{F}}\Lambda^{\frac12}\mat{V}^\dagger\qquad\text{where}\qquad\frac{1}{T}\widetilde{\mat{F}}^\dagger\widetilde{\mat{F}}=\mathds{1}_N.
    \end{equation}
    In order to re-conciliate this decomposition in terms of statistical uncorrelated modes $\widetilde{\mat{F}}$
    with the factors $\mat{F}$ of the model, the PCA solution \eqref{eq:PCA_sol} needs to be identified with Eq.~\eqref{eq:MODEL_mat}, which we 
    recopy here:
     \begin{equation}
     \Ret=\mat{F}\Wei+\mat{E}.
     \end{equation}
    
    The factors should explain as much as possible of the returns covariances (thus of the portfolio variance), 
    leaving only idiosyncratic residual volatility to be explained by the $e_i$'s.
    Said differently, only those eigenvalues having a significant amplitude
    should be kept in the identification of the spectral decomposition with the factor model.
    This procedure is known as ``eigenvalue clipping'' \citep{potters2009financial}. 
    Ordering the eigenvalues in decreasing order, and splitting the first $M$ 
    (subscript $_{M|}$) from the last $(N-M)$ (subscript $_{|N\!-\!M}$), it is straightforward to obtain the identification
    \begin{equation}\label{eq:X_PCA}
        {\Wei_{\text{PCA}}=\Lambda_{M|}^{\frac12}\mat{V}_{M|}^\dagger}.
    \end{equation}
    At this stage, the series of factors can be formally identified as the first $M$ spectral modes
    \begin{equation}\label{eq:id_f}
        \mat{F}=\widetilde{\mat{F}}_{M|}=(\Ret\mat{V}\Lambda^{-\frac12})_{M|}
    \end{equation}
    such that indeed $\frac{1}{T}\mat{F}^\dagger \mat{F}=\mathds{1}_M$.
    However, {the corresponding residuals are \emph{not} orthogonal}, since:
    \begin{equation}\label{eq:id_e}
        \mat{E}=\widetilde{\mat{F}}_{|N\!-\!M}\Lambda_{|N\!-\!M}^{\frac12}\mat{V}_{|N\!-\!M}^\dagger\quad\text{s.t.}
        \quad \frac{1}{T}\mat{E}^\dagger \mat{E}=\mat{V}_{|N\!-\!M}\Lambda_{|N\!-\!M}\mat{V}_{|N\!-\!M}^\dagger,
    \end{equation}
    and thus cannot be understood as idiosyncrasies of the returns time series.

%\subsubsection*{Factor weights}
The PCA can alternatively be thought of as the solution of 
\[
    \frac{1}{T}\Ret^\dagger \Ret=\Wei_{\text{PCA}}^\dagger \Wei_{\text{PCA}}\qquad\text{with}\qquad \Wei_{\text{PCA}}\Wei_{\text{PCA}}^\dagger \text{ diagonal,}
\]
where importantly $\Wei_{\text{PCA}}$ is full rank, but only the $M$ modes with largest amplitude $(\Wei_{\text{PCA}}\Wei_{\text{PCA}}^\dagger)_{kk}$, $k=1,\dots,M$ are kept 
\emph{after} the equation is solved. 

The factor model, on the other hand, can be seen as a close alternative to the ``eigenvalue clipping'' method. 
It rather attempts to minimize the distance between the off-diagonal elements of the LHS and the RHS with a matrix of weights $\Wei_{\text{F}}$ of restricted rank $M$:
\begin{equation}\label{eq:offdiag_content_apx}
    \argmin \left\|\frac{1}{T}\Ret^\dagger \Ret-\Wei_{\text{F}}^\dagger \Wei_{\text{F}}\right\|_{\text{off-diag}}.
\end{equation}
Numerically, we solve the above equation in the vicinity of the weights $\Wei_{\text{PCA}}$ corresponding to the $M$ largest principal components, 
and with a quadratic norm.
Notice that orthogonality of the lines of $\Wei_{\text{F}}$ obtained with this method is not granted,
as opposed to the PCA, but what matters is rather the fact that the factors are statistically uncorrelated.

How can one compare the information content of the factors on the one hand, and the Principal Components on the other? 
The idea is to measure the distance between the eigenspace spanned by the $M$ largest eigenvectors of the correlation matrix with
the $M$-dimensional eigenspace spanned by the weights $\Wei_{\text{F}}$ of the factor model. 
A natural measure for this distance was introduced 
by \cite{allez2011individual}, in terms of the $M \times M$ overlap matrix:
\begin{equation}\label{eq:defD}
D(M) = - \frac1M \ln \det \tilde\Wei_{\text{F}}^\dagger \Wei_{\text{PCA}}.
\end{equation}
where $\tilde\Wei_{\text{F}}$ is the orthonormalized set of vectors spanning the same subspace as $\Wei_{\text{F}}$. 
(Note that the $\Wei_{\text{PCA}}$ are by construction orthonormal: this is precisely why these weights are hard to interpret directly).

\begin{figure}
    \center
    \includegraphics[scale=.5]{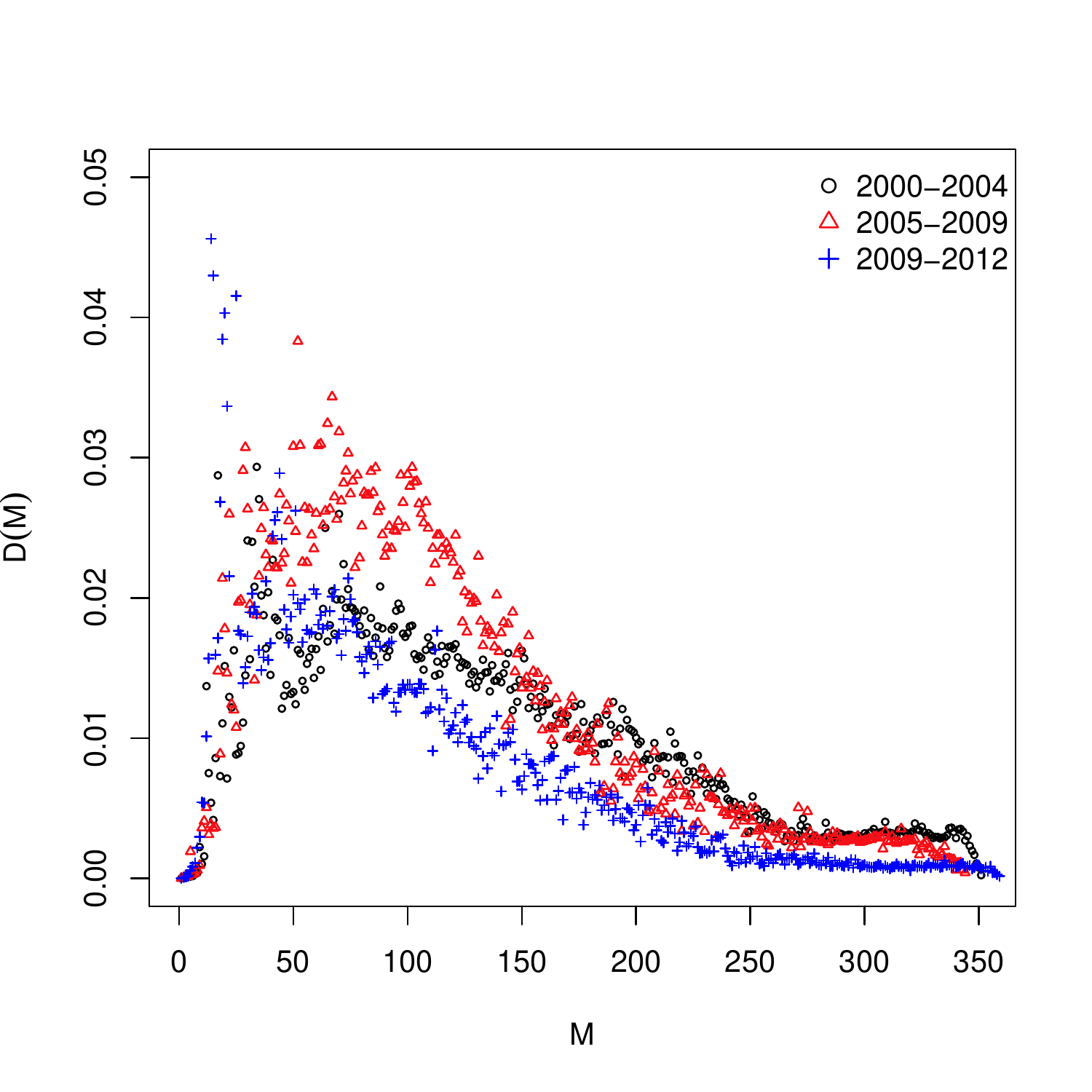}
    \caption{Overlap $D(M)$ between the spaces spanned by $\tilde\Wei_{\text{F}}$ and $\Wei_\text{PCA}$ when $M$ factors are retained, see definition in Eq.~\eqref{eq:defD}}
    \label{fig:genscalprod}
\end{figure}

The distance $D(M)$ for our data set is shown in Fig.~\ref{fig:genscalprod}. 
As expected, $D(M=1)$ is very small: both methods identify the most important ``mode'' or ``factor'' as the market itself. 
The overlap between $\Wei_{\text{F}}$ and $\Wei_{\text{PCA}}$ is in this case $99.9 \%$! 
Another trivial limit is $D(M=N)$, which is zero because the full space is by construction spanned in both cases. 
Between the two limits, we see that $D(M)$ always remains very small $< 4 \%$, 
which means that the information content is different, but similar for the two methods.%
\footnote{Some ``high'' values in the bulks may be due to arbitrary permutations in the labeling of the modes.
In fact, the PCA prior has a natural ordering (decreasing eigenvalues) but this order of relevant factors is not necessarily conserved
in the solution of Eq.~\eqref{eq:offdiag_content_apx}.}
Note that $D(M)$ is even less than $1 \%$ up to $M \approx 20$, i.e.\ for the most relevant eigenvalues. 

\section{Calibration of the dominant volatility model}\label{app:calibVol}

We introduce the Moment Generating Function (MGF) $M_{z}(p)\equiv\esp{\exp(p z)}$.
For the $\omega$'s and $\widetilde \omega$'s, that are assumed to be Gaussian, we have: $M_{\text{G}}(p)=\exp(s^2 p^2/2)$.
Expanding in powers of $p$, one gets
    \begin{align}\label{eq:MGF}
        M_{\Omega_0}(p) &=\Exp{\frac12 p^2+\frac{\zeta_0}{6}p^3+\frac{\kappa_0}{24}p^4}.
    \end{align}
It is then convenient to introduce the following ratio of MGF's:
    \begin{align}\label{eq:defPhi}
        \Phi_0(a,b) &=\frac{M_{\Omega_0}\!(a\!+\! b)}{M_{\Omega_0}\!(a)M_{\Omega_0}\!(b)},
    \end{align}
    as well as the Gaussian equivalent $\Phi_{\text{G}}(a,b)=\exp(ab)$.
    In logarithmic form, $\phi_0(a,b;p)=\frac{1}{p^2}\ln\!\Phi_0(pa,pb)$ is a polynomial in $p$ when expanding in cumulants.
    Indeed, with Eq.~\eqref{eq:MGF}, 
            \[\phi_0(a,b;p)=ab + \frac{p}{2}\zeta_0(a^2b+ab^2)+\frac{p^2}{12}\kappa_0(2a^3b+3a^2b^2+2ab^3),\]
            and $\phi_{\text{G}}(a,b;p)=ab$ is independent of $p$.
    Then the theoretical prediction for the matrix elements can be computed analytically:
    \begin{subequations}
    \label{eq:fkfleiej}
    \begin{align}
    \label{eq:fkfl0}
        \frac{1}{p^2}\ln\frac{\Esp{|f_k|^p|f_\ell|^p}}{\Esp{|f_k|^p}\Esp{|f_\ell|^p}}&=
        \phi_0(A_{k0},A_{\ell0};p)+\Big(\gamma(p) +s_{k}^2 \Big)\delta_{k\ell}\\
    \label{eq:fkel0}
        \frac{1}{p^2}\ln\frac{\Esp{|f_k|^p|e_i|^p}}{\Esp{|f_k|^p}\Esp{|e_i|^p}}&=
        \phi_0(A_{k0},B_{i0};p)\\
    \label{eq:ekel0}
        \frac{1}{p^2}\ln\frac{\Esp{|e_i|^p|e_j|^p}}{\Esp{|e_i|^p}\Esp{|e_j|^p}}&=
        \phi_0(B_{i0},B_{j0};p)+\Big(\gamma(p) +\widetilde s_{i}^2 \Big)\delta_{ij}
    \end{align}
    \end{subequations}
    where 
    \[
         \gamma(p)=\frac{1}{p^2}\ln\frac{\Esp{|\epsilon|^{2p}}}{\Esp{|\epsilon|^p}^2}
               =\frac{1}{p^2}\ln\!\left(\sqrt{\pi}\frac{\Gamma(\tfrac{1}{2}+p)}{\Gamma(\tfrac{1+p}{2})^2}\right)
    \]
    is the normalized $2p$-moment of the absolute value of Gaussian variables.
    
    For a Gaussian $\Omega_0$, the correlation matrices defined by Eqs.~\eqref{eq:fkfl0} and \eqref{eq:ekel0} would be trivially of rank 1, save the diagonal terms.
    If this was the case, the identification of $A_{\cdot0}$ and $B_{\cdot0}$ with the first eigenvectors of the corresponding matrices
    would be straightforward. Non-Gaussianity and specificities on the diagonal terms perturb this identification,
    but the overall picture is essentially the same story, as we show in Sect.~\ref{sec:modeling_vol} with the calibration results.
    
    The model estimation procedure is as follows (the linear weights $\Wei$ are previously estimated).
    As discussed above, there are $2(N+M+1)$ parameters to be estimated.
    Because the equations \eqref{eq:fkfleiej} are coupled through \eqref{eq:fkel0}, all parameters should in principle be estimated jointly.
    The corresponding optimization program would however be computer intensive, 
    and the stability of the solution would not be granted in such a large dimensional space.
    We proceed stepwise instead, by first estimating the parameters $A_{k0}, s_{k}, \zeta_0, \kappa_0$
    using the fac-fac predictions~\eqref{eq:fkfl0}, and then estimate the remaining parameters $B_{i0},\widetilde s_{i}$ from
    the res-res \emph{and} fac-res correlations for consistency. More precisely, our calibration procedure is as follows:
    
    \begin{enumerate}
        \item Estimate $A_{k0}$, $s_k$ and the non-Gaussianity parameters from Eq.~\eqref{eq:fkfl0}:
        \begin{equation}\label{eq:costfct_A}
            \min\left\{\sum_p\sum_{k,\ell}\left(
            \frac{1}{p^2}\ln\frac{\vev{|\mat{F}_{tk}\mat{F}_{t\ell}|^p}}{\vev{|\mat{F}_{tk}|^p}\vev{|\mat{F}_{t\ell}|^p}}
            -
            \frac{1}{p^2}\ln\frac{\Esp{|f_k|^p|f_\ell|^p}}{\Esp{|f_k|^p}\Esp{|f_\ell|^p}}
            \right)^2\right\}
        \end{equation}
        The sum on $p$ runs over eight values between $p=0.2$ and $p=2$ 
        and is crucial here to the estimation of the non-Gaussianity parameters, 
        since the loss function is independent of $p$ for Gaussian variables.
        This amounts to performing a best (joint!) quadratic fit of the curves similar to Fig.~\ref{fig:matelems}, 
        for each period.
        \item Estimate $B_{i0}$ from Eq.~\eqref{eq:fkel0}:
        \begin{subequations}\label{eq:loss_B}
        \begin{equation}
            \min\left\{\sum_{k,i}\left(
            \frac{1}{p^2}\ln\frac{\vev{|\mat{F}_{tk}\mat{E}_{ti}|^p}}{\vev{|\mat{F}_{tk}|^p}\vev{|\mat{E}_{ti}|^p}}
            -
            \frac{1}{p^2}\ln\frac{\Esp{|f_k|^p|e_i|^p}}{\Esp{|f_k|^p}\Esp{|e_i|^p}}
            \right)^2\right\}
        \end{equation}
        or jointly with $\widetilde s_{i}$ from Eq.~\eqref{eq:ekel0}, as the vector solution of
        \begin{equation}
            \min\left\{\sum_{i,j}\left(
            \frac{1}{p^2}\ln\frac{\vev{|\mat{E}_{ti}\mat{E}_{tj}|^p}}{\vev{|\mat{E}_{ti}|^p}\vev{|\mat{E}_{tj}|^p}}
            -
            \frac{1}{p^2}\ln\frac{\Esp{|e_i|^p|e_j|^p}}{\Esp{|e_i|^p}\Esp{|e_j|^p}}
            \right)^2\right\}
        \end{equation}
        \end{subequations}
        (here it is too intensive to calculate the optimum in the $N$-dimensional 
        space for all values of $p$ so we take a single value, 
        typically $p=1$ if we intend to reproduce best absolute correlations, 
               or $p=2$ if we favor quadratic correlations).
        
    \end{enumerate}
    The convergence is ensured by starting close to the solution, namely taking as prior the first eigenvector of the corresponding matrices.

\section{Test of the out-of-sample performance of correlation models}\label{app:ISOS}
    
    The protocol proposed by \cite{potters2009financial} to compare the out-of-sample risk of different correlation models is as follows:
    
    \begin{enumerate}
    \item The model is calibrated in windows of $T^{\scriptscriptstyle \text{IS}}=2N=524$ days.
          An optimal portfolio is built and a corresponding risk measure is computed over the window used for estimation:
          this is the In-sample risk. 
          We consider below two kinds of risks corresponding to two different portfolios:
          (i)  the quadratic risk of a basket of returns, that will assess the quality of the \emph{linear} elements of the model; and
          (ii) the quadratic risk of a basket of (centered and normalized) absolute returns, that will assess the quality of the \emph{volatility} description of the model.
    \item The same risk measures are computed Out-of-sample on a small period of $T^{\scriptscriptstyle \text{OS}}=59$ days (three months) following the estimation period.
    \item The sliding lags are chosen so that the control samples are non-overlapping,
          i.e.\ at dates $\tau=T^{\scriptscriptstyle \text{IS}}+n\times T^{\scriptscriptstyle \text{OS}}+1$, $n=0,1,2,\ldots$.
          Sliding windows will be indexed with parenthesis notation ``$(\tau)$'', 
          in order to avoid confusion with regular time stamps $t$ of the running dates.
    \end{enumerate}
    
    We then build a portfolio of assets $y_i$ knowing their historical time series $\mat{Y}_{ti}$, which can be returns in the standard case, but also 
    absolute returns or squared returns when one has non-linear assets in mind (such as options, for example).
    
    For a given covariance matrix $\rho_{\ij}=\operatorname{cov}{(y_i,y_j)}$, 
    optimal portfolio weights can be computed in the sense of Markowitz:
    \begin{equation}\label{eq:markowitz_w}
                    \vect{w}^*(\tau) =\frac{\rho^{-1}\vect{g}(\tau)}{\vect{g}(\tau)^\dagger\rho^{-1}\vect{g}(\tau)}
    \end{equation}
    where we consider an omniscient stationary predictor of returns
    \begin{equation}\label{eq:markowitz_g}
    g_i(\tau) =\frac{\mat{Y}_{\tau i}}{\sqrt{\frac{1}{N}\sum_j \mat{Y}_{\tau j}^2}}
    \end{equation}
    and a unit total gain $\mathcal{G}\equiv\vect{g}^\dagger\vect{w}^*=1$. 
    This means that the in-sample/out-of-sample test procedure applied below is intended to measure only {risk}
    and not the risk-return trade-off (Sharp ratio) as is usual e.g.\ when back-testing financial strategies.
    Indeed what we ultimately want to conclude is whether our model of stock returns 
    allows to have a better view of dependences and thus to better diversify away the risk 
    (since we work with normalized returns, we are not concerned with individual variances but only care for dependences).
    
    Quadratic risk is essentially a measure of expected small fluctuations of the portfolio value:
    \begin{equation}\label{eq:def_risk}
        \mathcal{R}^2(\tau)=\frac{1}{T'}\sum_{t'}\frac{1}{N}\sum_{i=1}^N\left[\mat{Y}_{t'i}\frac{w_i^*(\tau)}{\sigma_i^{\scriptscriptstyle \text{IS}}(\tau)}\right]^2
    \end{equation}
    where, for convenience, the asset returns are normalized by a rolling in-sample estimate of their volatility
    $\sigma^{\scriptscriptstyle \text{IS}}(\tau)$ 
    --- although the returns have been normalized over the whole period, they may not be close to unit-variance in-sample
    because of low-frequency regime switches in the volatility.
    This risk is computed both 
             in-sample (in which case $T'=T^{\scriptscriptstyle \text{IS}}$ and $t'=\tau-T', \ldots, \tau-1 $) 
     and out of sample ($T'=T^{\scriptscriptstyle \text{OS}}$ and $t'=\tau+1 , \ldots, \tau+T'$),
     for different input correlation matrices in Eq.~\eqref{eq:markowitz_w}.

\bibliographystyle{QF/rQUF}
\bibliography{../biblio_all}

%%%%%%%%%%%%%%%%%%%%%%%%%%%%%%%%%%%%

\end{document}